\documentclass[traditabstract]{aa}

\usepackage{aecompl}

\usepackage{array}
\usepackage{amsmath}
\usepackage{graphicx}
\usepackage{amssymb}
\usepackage[round]{natbib}

\usepackage{calc}
\usepackage{subfigure}

\usepackage[colorinlistoftodos]{todonotes}

\usepackage{graphicx,bm}
\usepackage[colorlinks,citecolor=blue]{hyperref}
\usepackage{placeins}

\providecommand{\aap}{Astron.\ Astrophys.}
\providecommand{\apjs}{Astrophys.\ J.\ Suppl.}
\providecommand{\apjl}{Astrophys.\ J.}

\begin{document}

\title{Neutron stars in accreting systems -- Signatures of the QCD phase transition}

\author{
Noshad~Khosravi Largani\inst{1}\thanks{{\em email:} noshad.khosravilargani@uwr.edu.pl}
\and Tobias~Fischer\inst{1}\thanks{{\em email:} tobias.fischer@uwr.edu.pl}
\and Shota~Shibagaki\inst{2,3,4}
\and Pablo~Cerd{\'a}-Dur{\'a}n\inst{5,6}
\and Alejandro~Torres-Forn{\'e}\inst{5,6}
}

\institute{
Institute for Theoretical Physics, University of Wroc{\l}aw, Plac Maksa Borna 9, 50-204 Wroc{\l}aw, Poland
\and Incubator of Scientific Excellence---Centre for Simulations of Superdense Fluids, University of Wroc{\l}aw, Plac Maksa Borna 9, 50-204 Wroc{\l}aw, Poland
\and Helmholtz-Zentrum Dresden-Rossendorf (HZDR), Bautzner Landstrasse 400, 01328 Dresden, Germany
\and Center for Advanced Systems Understanding (CASUS), Untermarkt 20, 02826 G\"orlitz, Germany
\and Departament d’Astronomia i Astrof{\'i}sica, Universitat de Val{\'e}ncia, C/ Dr Moliner 50, 46100, Burjassot (Val{\'e}ncia), Spain
\and Observatori Astronòmic, Universitat de València, E-46980, Paterna (València), Spain
}

\authorrunning{N.~Khosravi Largani et al.}

\date{Received date / Accepted date }

\abstract{
Neutron stars (NS) that are born in binary systems with a main-sequence star companion can experience mass transfer, resulting in the accumulation of material at the surface of the NS.
This, in turn, leads to the continuous growth of the NS mass and the associated steepening of the gravitational potential. 
Supposing the central density surpasses the onset for the phase transition from nuclear, generally hadronic matter to deconfined quark-gluon plasma, which is a quantity currently constrained solely from an upper limit by asymptotic freedom in quantum chromodynamics (QCD), the system may experience a dynamic response due to the appearance of additional degrees of freedom in the equation of state (EOS).
This dynamical response might give rise to a rapid softening of the EOS during the transition in the hadron-quark matter co-existence region. 
While this phenomenon has long been studied in the context of hydrostatic configurations, the dynamical implications of this problem are still incompletely understood. 
It is the purpose of the present paper to simulate the dynamics of NSs with previously accreted envelopes caused by the presence of a first-order QCD phase transition. 
Therefore, we employed the neutrino radiation hydrodynamics treatment based on the fully general relativistic approach in spherical symmetry, implementing a three-flavor Boltzmann neutrino transport and a microscopic model EOS that contains a first-order hadron-quark phase transition. 
The associated neutrino signal shows a sudden rise in the neutrino fluxes and average energies, becoming observable for the present generation of neutrino detectors for a galactic event, and a gravitational wave mode analysis revealed the behaviors of the dominant $f$ mode and the first and the second gravity $g$ modes that are excited during the NS evolution across the QCD phase transition.}


\maketitle

\section{Introduction}
\label{sec:intro}

Accreting neutron stars (NSs) in binary systems have long been considered subjects of gravitational wave (GW) emission \citep[c.f.][]{Wagoner1984ApJ278}, such as through ellipticity and internal oscillations, which has been studied extensively in the context of low-mass X-ray binary systems by \citet{Watts2008MNRAS389} with potential future GW detection prospects \citep[see also][]{Andersson2011GReGr43}. Unstable oscillation modes have been studied, too, in the context of accreting NSs \citep[c.f.][and references therein]{Lasky2015PASA32} as well as GW emission due to star quakes \citep[see][and references therein]{Ruderman1969Natur223,Giliberti2022MNRAS511} and due to high magnetic fields \citep[c.f.][and references therein]{Bonazzola1996A&A312}. (For recent works about magnetically deformed NSs, see \citet{Haskell2008MNRAS385} and \citet{Haskell2020MNRAS493}.) 
One of the largest uncertainties in all of these considerations is the still incompletely understood equation of state (EOS) concerning the NS crust, especially the stability of the crust \citep[][]{ChamelHaensel2008LRR11} and the high-density phase in excess of nuclear saturation density. In particular, the latter concerns the presence of hadronic resonances as well as a transition to the quark-gluon plasma. From numerical solutions of quantum chromodynamics (QCD), which is the theory that describes strong interactions with quarks and gluons as fundamental degrees of freedom, a crossover transition occurs at a pseudocritical temperature in the range of 150 to 160~MeV and at a vanishing baryon density \citep[c.f.][]{Bazavov:2019}. Even though perturbative QCD has been probed to potentially constrain the high-density EOS of compact stars \citep[c.f.][and references therein]{Kurkela:2014vha,Annala20}, the conditions encountered at the NS interior correspond to the non-perturbative regime of QCD, and hence phenomenological quark matter models have commonly been employed in astrophysical studies. These models assume a first-order transition with a phase transition construction from a given hadronic model EOS \citep[][]{glendenning2012compact}. 
Neutron star glitches, such as those observed in \citet{Espinoza2011MNRAS414_glitches_obs}, speculated to be related to possible phase transitions at the NS interior due to the appearance of deconfined QCD degrees of freedom as well as a transition to a state of superfluidity have been investigated \citep[for classical works, see][and references therein]{Anderson1975Natur256_glitches,Alpar1984ApJ282_glitches}. (For a comprehensive review of the possible mechanisms for glitches, see \citet{Haskell2015IJMPD_glitches-mechanism_review}.)

Complementary, methods of NS asteroseismology have long been explored in the context of compact stars. The linear perturbation theory was implemented in \citet{Andersson1998MNRAS299}, and references therein, to predict possible GW emission characteristics. Such an approach has been demonstrated to provide reliable frequency estimates for the dominant modes, known as $f$ mode, as well as for gravity $g_l$ modes and pressure $p_l$ modes. While the $f$ mode has been referred to as the damping of oscillations \citep[c.f.][]{Benhar2004PhRvD70}, the presence of $g$ (transverse) and $p$ (longitudinal) modes are related to buoyancy and sound waves, respectively. Furthermore, astroseismology has been proposed as a tool to probe the stellar properties, which implies the EOS. Astroseismology investigations of NS were conducted by \citet{Sotani2012PhRvL108} , while proto-NSs were investigated by \citet{Sotani2016PhRvD94}, \citet{sotani2017PhRvD96}, \citet{Torres-Forne2018MNRAS474}, and \cite{Torres-Forne2019MNRAS482} as well as by \citet{sotani2019PhRvD100} toward black hole formation in failed core-collapse supernovae \citep[see also][]{CerdaDuran13}. In addition, various universal relations of NS bulk properties have been studied to remove the EOS dependence, for example, those relating the maximum mass and corresponding radii of rapidly and non-rotating configurations as well as the moment of inertia and NS tidal deformability \citep[for a recent review, see][and references therein]{YagiYunes2017PhyRep}. These investigations were extended to include finite temperature effects \citep[c.f.][]{Raduta2020MNRAS499,Khosravi2022MNRAS515}. Universal relations have also been suggested to exist for several GW modes, as deduced from detailed astroseismology analyses in the content of NS in \citet{Sotani2021PhRvD104a_universal-relations_NS} and for proto-NS by \citet{Torres-Forne12019PhRvL123}, \citet{Warren2020}, \citet{Sotani2021PhRvD104b_unversal-relations_SN}, and \cite{Mori2023}. These studies, despite some tension about the exact dependencies, such as surface gravity and mean density, will eventually reveal details about the EOS from future GW detections. 

A detailed analysis within the context of QCD phase transitions at the NS interior has been considered within general relativistic hydrodynamics simulations of isolated NSs in axial symmetry by \citet{Dimmelmeier09}, where the phase transition occurs due to the loss of angular momentum, and in hydrostatic models by \citet{Sotani2023PhRvD108_universal-relations_QCD}, who probed the high-density EOS. The latter study found that the configurations with quark matter at their interior have generally lower dominant $f$-mode frequencies by several 100~Hz, compared to the hadronic reference case, depending on the compact star's mass. On the other hand, hydrodynamical simulations featuring a first-order QCD phase transition have been performed extensively in the context of core-collapse supernovae \citep[][]{Sagert09,Fischer11,Fischer18} and binary NS mergers \citep[][]{Bauswein19,Most19}. In these studies, special emphasis was placed on predictions for potentially observable multi-messenger signals in neutrino and GW emissions \citep[see also][]{Zha20,Kuroda2022} as well as constraining bulk properties from a potential future observation of the core-collapse supernova neutrino signal in \citet{Khosravi2023arXiv230412316K} and the binary NS merger GW signals in \citet{Blacker20}.  \citet{Kuroda2022} found loud core-collapse supernova GW signals from the first-order QCD phase transition, with frequencies in the range of several kilohertz and amplitudes in excess of the ordinary neutrino-driven supernova explosions and even of black hole formation. 

In this article, we extend the study of \citet{Sotani2023PhRvD108_universal-relations_QCD} by performing neutrino radiation hydrodynamics simulations of NSs in accreting systems as an alternative to the astrophysical scenario in order to probe a possible QCD phase transition. We implemented a well-calibrated nuclear matter EOS from \citet{Hempel2010NuPhA837} based on the DD2F relativistic mean field parametrization of \citet{Typel10} with density-dependent nucleon-meson couplings together with a first-order phase transition to the quark-gluon plasma based on the relativistic density function EOS of \citet{Bastian2021PhRvD103}. The resulting hybrid EOS is consistent with the present nuclear physics constraints such as nuclear saturation properties, the supersaturation density constraint derived from the elliptic flow analysis of heavy-ion collision experiments by \citet{Danielewicz:2002}, and the astrophysical constraints derived from observations of pulsar timing. 

From population synthesis, it is presently understood that about 30-50\% of massive stars are in binary systems \citep[c.f.][and references therein]{Eldridge2017hsnbook}, some of which result in high-mass X-ray binaries  \citep[c.f.][and references therein]{TaurisHeuvel2006csxs.book..623T}. 
Even though stellar evolution predicts potentially very different tracks for single and binary star systems \citep[c.f.][]{Sana2012Sci337}, it is natural to assume that some of the binary systems will end up having an NS, that is, after the first core-collapse supernova explosion has taken place, and a main-sequence companion star. The latter can well be a low- or intermediate-mass star. The lifetime of such systems depends on their properties, including age, metallicity, and geometry, such as the spatial separation and angular velocities, as well as those of the companion star (e.g., its mass). Other impacts on the lifetime can come from whether the binary system's configuration enables mass transfer from the companion star onto the NS and whether a common envelope phase occurs \citep[][]{Keegans2019}, which is enabled when Roche-Lobe overflow occurs \citep[c.f.][and references therein]{Tauris2013ApJ778_ultra-stripped}. 
The details leading to accreting NS systems are left aside in this present investigation. Instead, we considered that an NS has undergone mass transfer for a certain period and further that mass accretion has ceased. This is similar to what has been considered in accreting white dwarf systems leading to novae \citep[][]{Metzger2021ARA&A59_ReviewNovae}. Furthermore, it is expected that mass transfer from the companion star heats the NS's surface and, to a lesser extent, the NS interior. In excess of temperatures of several $10^9$~K ($T=6\times 10^9$~K corresponding to $0.517$~MeV), neutrinos can be produced from a variety of weak reactions. Taking those into account in the simulations featuring a three-flavor Boltzmann neutrino transport allows for the prediction of not only the neutrino fluxes and their spectra but also the study of the potential impact on mass ejection associated with the NS evolution. 

The manuscript is organized as follows. In Sec.~\ref{sec:setup} the simulation setup is introduced followed by the discussion of the neutrino radiation hydrodynamics simulations in Sec.~\ref{sec:simulations}, including the neutrino emission. Section~\ref{sec:modes} presents the GW mode analysis. The paper closes with a summary in Sec.~\ref{sec:summary}.

\section{Equation of state and initial conditions}
\label{sec:setup}
\begin{figure}[t!]
\begin{center}
\includegraphics[angle=0.,width=1\columnwidth]{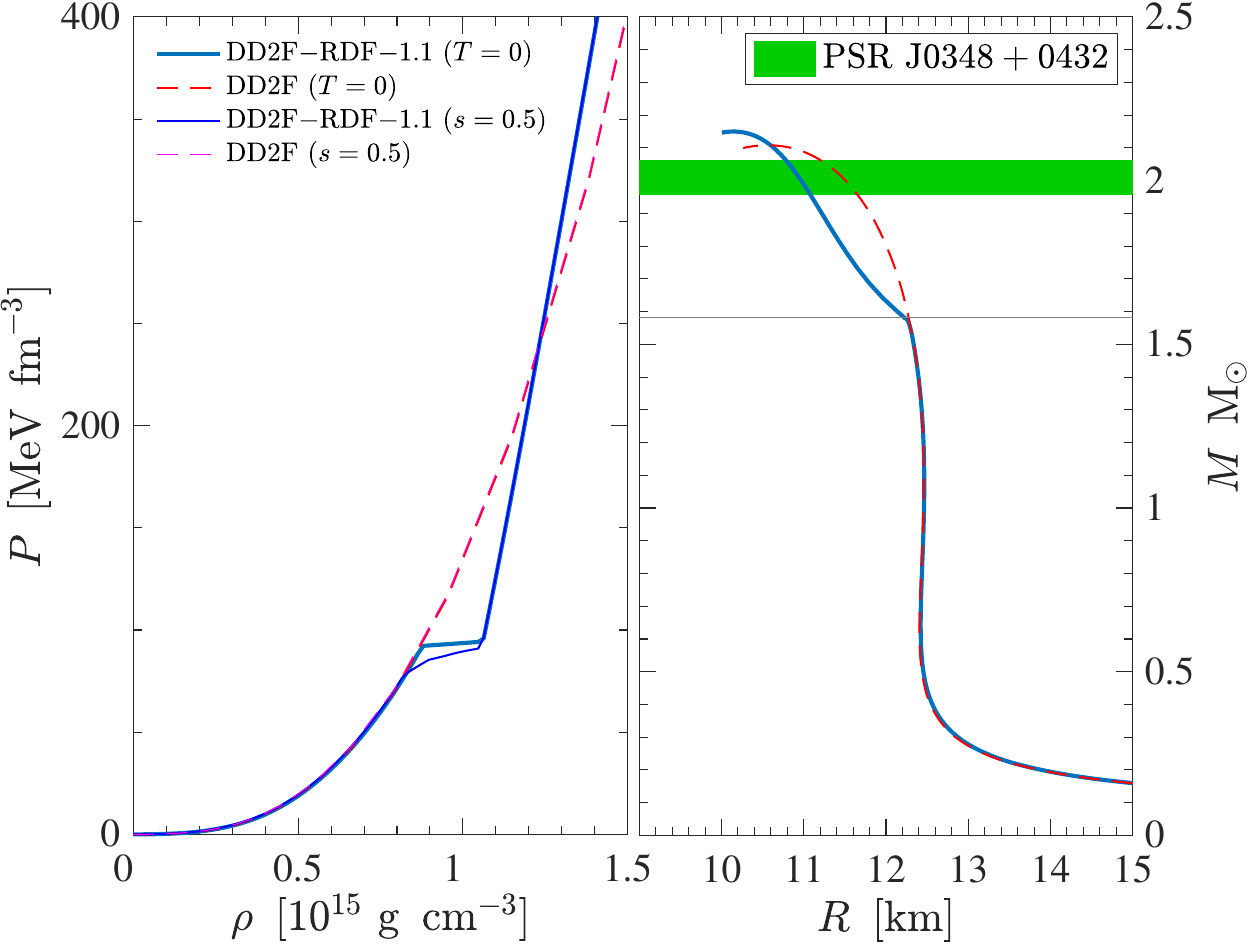}
\caption{~Comparison of the EOS at $\beta$-equilibrium for zero temperature $(T=0)$ and for a constant entropy per particle of $s=0.5~k_{\rm B}$ for DD2F-RDF-1.1 (solid light and dark blue lines) and DD2F (dashed red and magenta lines) with the corresponding DD2F hadronic EOS (red and magenta dashed lines, respectively, which are on top of each other for both configurations).
{\em Left panel:}~Pressure $P$ versus rest-mass density $\rho$. 
{\em Right panel:}~Mass $M$ radius $R$ relation. Included is an example for the high-precision maximum mass constraint of $2.01\pm0.04$~M$_\odot$ (light grey band) deduced from the radio observations of PSR~J0348+0432 of \citet{Antoniadis13}. 
The horizontal grey line in the mass-radius diagram marks the onset mass for the quark-hadron phase transition at $M=1.582$~M$_\odot$ for the EOS with $s=0.5~k_{\rm B}$.}
\label{fig:eos_MR}
\end{center}
\end{figure}

To construct the initial conditions for the numerical study of systems that are composed of NSs with previously accreted envelopes, we employed the hadronic EOS catalog of \citet{Hempel12} together with the relativistic density functional (RDF) quark-matter hadron EOS catalog of \citet{Bastian2021PhRvD103}, both of which are available as multi-purpose tabulations for astrophysical application at the CompStar Online Supernovae Equations of State (CompOSE) EOS database.\footnote{The EOS data and routines provided by CompOSE can be downloaded from the web site
https://compose.obspm.fr} 
For the hadronic EOS, the DD2F relativistic mean field (RMF) model of \citet{Typel10} was selected, and for the quark-hadron EOS, the DD2F-RDF-1.1 parametrization was employed as representative for the entire class of RDF models. At densities below normal nuclear matter density\footnote{The saturation density for DD2F is 0.145~fm$^{-3}$, or equivalently $2.44\times 10^{14}$~g~cm$^{-3}$} and low temperatures, the RMF EOS was extended into the modified nuclear statistical equilibrium (NSE) model of \citet{Hempel2010NuPhA837} with several thousand nuclear species \citep[for a review of this class of hadronic EOS in supernova simulations, see][and references therein]{Fischer17}. Figure~\ref{fig:eos_MR} (left panel) compares these two EOSs at the condition of $\beta$-equilibrium and for two constant values of the entropy per baryon of $s=0.1~k_{\rm B}$ and $s=0.5~k_{\rm B}$, neglecting contributions from neutrinos (i.e., the electron lepton number equals the electron abundance, $Y_{\rm L}=Y_e$). This assumption is justified, as neutrinos play no role in the stability of NSs for the entropies explored in the present study.

The DD2F-RDF-1.1 hybrid EOS of \citet{Bastian2021PhRvD103} implements a first-order hadron-quark phase transition from the DD2F hadronic EOS. At the present conditions of $s=0.5~k_{\rm B}$, we obtained an onset density for the phase transition of $\rho=8.8\times 10^{14}$~g~cm$^{-3}$ and an onset mass of 1.582~M$_\odot$, as is illustrated via the horizontal grey line in the right panel of Fig.~\ref{fig:eos_MR}. We note that the EOS bulk property differences for different entropies in the range of $s=0.1$-$0.5~k_{\rm B}$ and $T=0$ configurations are on the order of one percent \citep[see Table~I in][]{Bastian2021PhRvD103}. Larger differences arise in the phase transition region, as is illustrated in the left panel in Fig.~\ref{fig:eos_MR}, which shows a comparison of the $T=0$ and $s=0.5~k_{\rm B}$ cases. The onset density for the phase transition decreases and the change of the pressure slope is less abrupt for the finite entropy situation. This feature of the RDF class of hadron-quark matter hybrid EOS has been discussed in detail in the example of core-collapse supernova simulations in \citet{Fischer18}, \citet{Fischer:2021}, and recently in \citet{Khosravi2023arXiv230412316K} as well as for (non)rotating hot and cold NS in \citet{Khosravi2022MNRAS515}. However, for somewhat larger entropies of  $s=3~k_{\rm B}$, it leaves a stronger impact on the hybrid (proto)NS maximum mass and radius. 

\begin{figure*}[t!]
\begin{center}
\includegraphics[angle=0.,width=1.7\columnwidth]{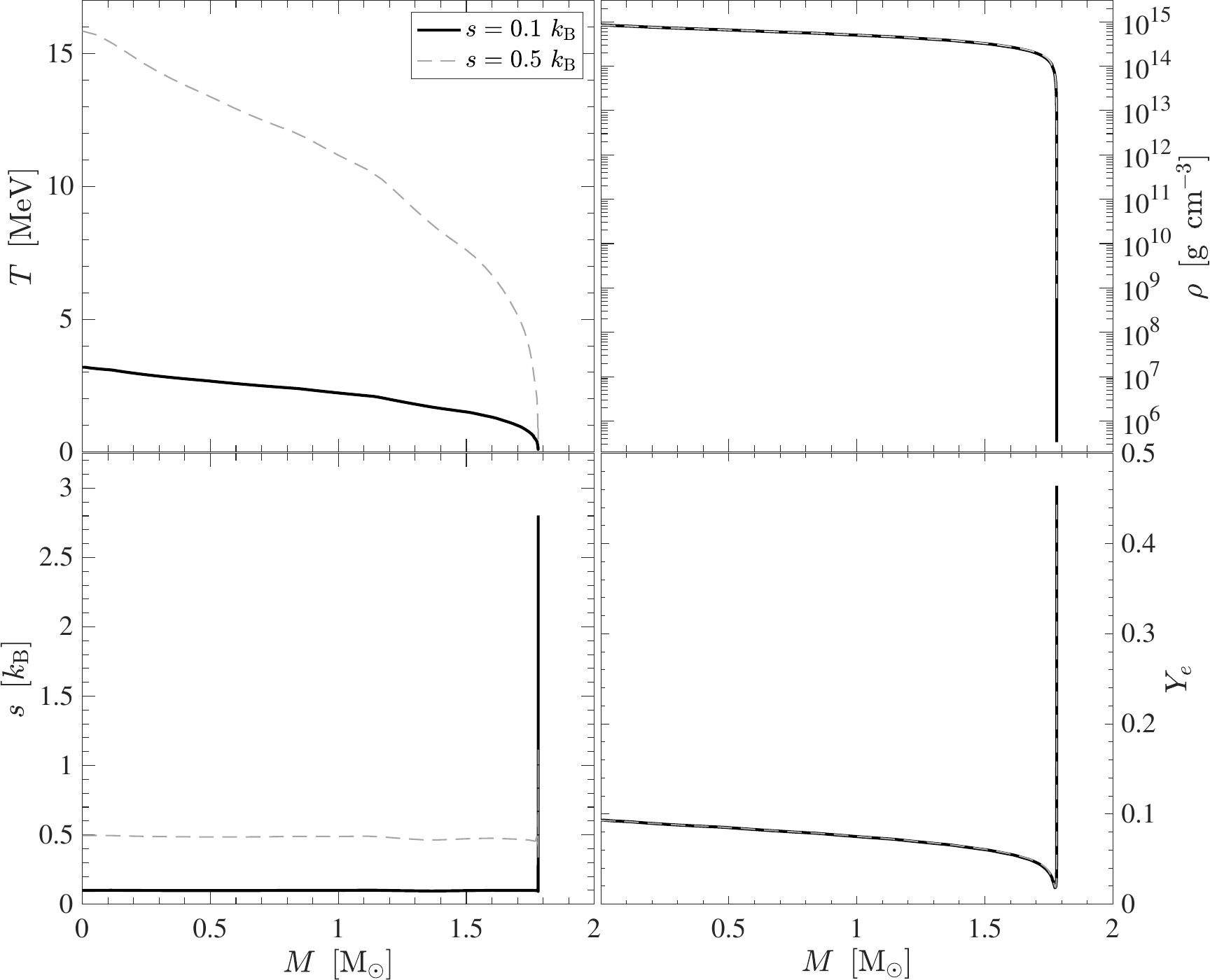}
\caption{~Initial conditions for the NS simulations based on the DD2F-RDF-1.1 hadron-quark hybrid EOS for an NS with total baryon mass of $M=1.781$~M$_\odot$ (corresponding to a gravitational mass of $M_{\rm G}=1.588$~M$_\odot$). The temperature $T$ is shown in the upper-left panel, rest-mass density $\rho$ is in the upper-right panel, entropy per baryon $s$ is shown in the lower-left panel, and electron abundance $Y_e$ is in the lower-right panel, as a function of the enclosed baryon mass. Two configurations with a constant central entropy of $s=0.1~k_{\rm B}$ (black solid lines) and $s=0.5~k_{\rm B}$ (grey dashed lines) are compared.}
\label{fig:initial}
\end{center}
\end{figure*}

The quark matter phase of the RDF EOS is based on the concept of the string-flip model, which was originally developed by \citet{Goddard1973NuPhB56_stringflip} and \citet{JohnsonThorn1976PhRvD13_stringflip} and later extended by \citet{Ropke86_stringflip} and \citet{Horowitz1985PhRvD31_stringflip}. The model mimics confinement through divergent medium-dependent quark masses, which has been extended recently for applications to hydrodynamical simulations of heavy-ion collisions in \citet{Bastian2023NuPhA103}. Of particular relevance for astrophysics are repulsive interactions, implemented here through leading and higher-order vector interaction channels following \citet{Benic:2015} and \citet{Kaltenborn17}, similar to the case in the simplistic quark-matter model EOS of the Nambu--Jona-Lasinio type \citep{NJL:1961} as well as the vector-interaction enhanced bag models of \citet{Klaehn:2015}. Repulsive interactions are essential to obtaining the maximum NS masses---hybrid stars if quark matter is present at their interior---consistent with the current constraints of the maximum mass of about 2~M$_\odot$ deduced from high-precision radio observations \citep[c.f.][]{Antoniadis13,Fonseca:2021}. The DD2F-RDF-1.1 hybrid EOS features a maximum mass of 2.150~M$_\odot$ for $s=0.5~k_{\rm B}$ and 2.155~M$_\odot$ for $T=0$ \citep[for a detailed discussion about the differences of finite entropy per particle and $T=0$ NS and hybrid stars, see][]{Khosravi2022MNRAS515}. 

Precision NS radius constraints are available since the observation analysis of quiescent low-mass X-ray binaries by \citet{Steiner2010ApJ722}, with $R=11$-$12$~km for intermediate-mass NSs of $M\simeq 1.4$~M$_\odot$. These constraints have been further strengthened by the analysis of the GW data from the binary NS merger event GW170817 provided by \citet{Abbott2018PhRvL121_NSEOS}, which was reanalyzed by \citet{Lattimer2018PhRvL121_NSEOS} while also taking into account the binary NS mass ratio. Further, high-precision NS radii constraints are available from the Neutron star Interior Composition Explorer (NICER) NASA mission, which are consistent with the present DD2F-RDF-1.1 EOS. The radii of massive NSs of around 2.0~M$_\odot$ is $R_{2.0}=11.1$~km, in agreement with the values derived by \citet{NICER_Riley2021} and \citet{NICER_Miller2021}, and the radii of intermediate-mass NSs of around 1.5~M$_\odot$ is $R_{1.5}=12.4$~km, in agreement with the analysis of \citet{NICER_Watts2019} and \citet{NICER_Miller2019}.

The initial conditions were constructed for the hydrodynamical simulations of NSs with an accreted envelope. Therefore, hydrostatic NS solutions based on the Tolman-Oppenheimer-Volkoff (TOV) equations were computed for an NS of selected baryon mass of $M=1.781$~M$_\odot$ for the hadronic DD2F EOS. It corresponds to a gravitational mass of $M_{\rm G}=1.588$~M$_\odot$. 
 To this end, we constructed a finite temperature EOS at $\beta$-equilibrium and implemented a numerical root-finding algorithm to yield finite entropy per baryon configurations. Two of these are shown in Fig.~\ref{fig:eos_MR} for $s=0.1~k_{\rm B}$ and compared with a configuration of $s=0.5~k_{\rm B}$ that neglects the contributions from muons and neutrinos (i.e., $Y_\mu=0$ and $Y_{\rm L}\equiv Y_e$, respectively). The TOV integration was then performed for the aforementioned enclosed baryon mass, giving rise to the one-dimensional profiles of rest-mass density, temperature, and $Y_e$ shown in Fig.~\ref{fig:initial} as a function of the enclosed baryon mass $M$. 

It became evident that the structure of the NS is independent of all the thermodynamic quantities, such as the rest-mass density and $Y_e$, for this entropy range except central temperatures, which vary between about 3 and 16~MeV, respectively, (upper-left panel in Fig.~\ref{fig:initial}) for the entropies per baryon explored here of $s=0.1~k_{\rm B}$ and $s=0.5~k_{\rm B}$. 
However, the thermal contributions to the high-density EOS are comparable for this range of entropies such that both configurations have the same central densities of $\rho=8.5\times 10^{14}$~g~cm$^{-3}$ (upper-right panel in Fig.~\ref{fig:eos_MR}). 
Toward the star's surface, the temperature decreases to around $T=0.5$~MeV, and the electron abundance (lower-right panel) increases simultaneously, from the minimum value of $Y_e=0.018$ up to values of $Y_e=0.4642$ at the lowest density considered in this study. 
For simplicity, low-density matter was assumed to be in nuclear statistical equilibrium (NSE) in the hydrodynamical simulations (i.e., the NS envelope and crust).
According to the modified NSE EOS of \citet{Hempel2010NuPhA837}, the nuclear composition features $^{56}$Ni, the most stable nucleus with the largest nuclear binding energy per baryon. 
Following \citet{Hempel2010NuPhA837} and \citet{Hempel12}, the NS crust-core transition is implemented via the excluded volume approach, therefore constructing a first-order phase transition that ensures the disappearance of all nuclei at the nuclear saturation density. A comparison of this approach with other microscopic nuclear calculations has been discussed in \citet{Fischer2020PhRvC102}.

We assumed an accreted envelope that contains $M_{\rm envelope}=0.018$~M$_\odot$. 
We matched it in an ad hoc manner to the TOV profiles of the hydrostatic and isentropic NS interior and crust solutions, assuming the modified NSE of \citet{Hempel2010NuPhA837} for a constant temperature of 0.45~MeV, with $^{56}$Ni as the nuclear composition. The matching procedure implemented ensures a smooth density extension toward lower densities with increasing radii.  

This matching was constructed such that the resulting NS, with a total mass of $M_{\rm G}=1.588$~M$_\odot$, belongs only to the hybrid branch in the mass-radius diagram. We note that the onset mass is 1.57~M$_\odot$ for the DD2F-RDF-1.1 EOS \citep[see Table~I in][]{Bastian2021PhRvD103}. The continuous mass transfer from the secondary companion star is expected to gradually heat the stellar interior, as the gravitational potential steepens continuously during this process, for which we selected the aforementioned values of central entropy. However, the entropy of the outer parts of the envelope rises sharply at low density, reaching values up to $s=1$-$3~k_{\rm B}$ (see Fig.~\ref{fig:eos_MR}), which is due to the assumed constant temperature of 0.45~MeV being dominated by radiation.

We note that the heated envelope might be subject to explosive hydrogen burning processes and the subsequent emission of type-I X-ray bursts, which are driven by charged particle reactions and yield the production of isotopes up to a nuclear mass number on the order of $A\simeq 100$ \citep[c.f.][and references therein]{WallaceWoosley1981ApJS45,JossRappaport1984ARA&A22,Lewin1993SSRv62,Schatz1999ApJ524,Fisker2008ApJS174}. This is expected to result in the continuous growth of the total enclosed mass on a longer timescale due to mass accretion rates on the order of $\dot{M}\gtrsim 10^{-10}-10^{-8}$~M$_\odot$ per year---depending on the geometry and other properties of the binary system---since only a certain fraction of the previously accreted envelope would contribute to the explosive hydrogen burning. The remaining material will fall back and settle at the NS crust, adding up to its gravitational and baryonic mass. It is this process that leads to the dynamical features of a QCD phase transition when reaching the critical mass for the onset of quark matter, which we discuss in the following section.

\section{Neutron star simulations with phase transition}
\label{sec:simulations}
We note that the potential occurrence of explosive hydrogen burning is neglected in this work through the assumption of an NSE envelope since our focus is on the dynamical implications of a first-order QCD phase transition at the NS interior.
The initial conditions constructed in Sec.~\ref{sec:setup} corresponding to the NS with an accreted envelope are at the boundary between the hadronic phase and the onset conditions for the hadron-quark phase transition. In fact, only the envelope mass of 0.0181~M$_\odot$ is above the mass onset. These solutions of the hydrostatic equations for $s=0.1~k_{\rm B}$ are implemented into the neutrino radiation hydrodynamics model as initial conditions, considering the impact of the hadron-quark phase transition. In Appendix~\ref{sec:appendix1}, we confirm the hydrostatic equilibrium solution obtained within the hydrodynamics approach based on the DD2F EOS, implementing a high radial resolution of 300 mass mesh points. 

\begin{figure}[t!]
\centering
\includegraphics[angle=0.,width=1.\columnwidth]{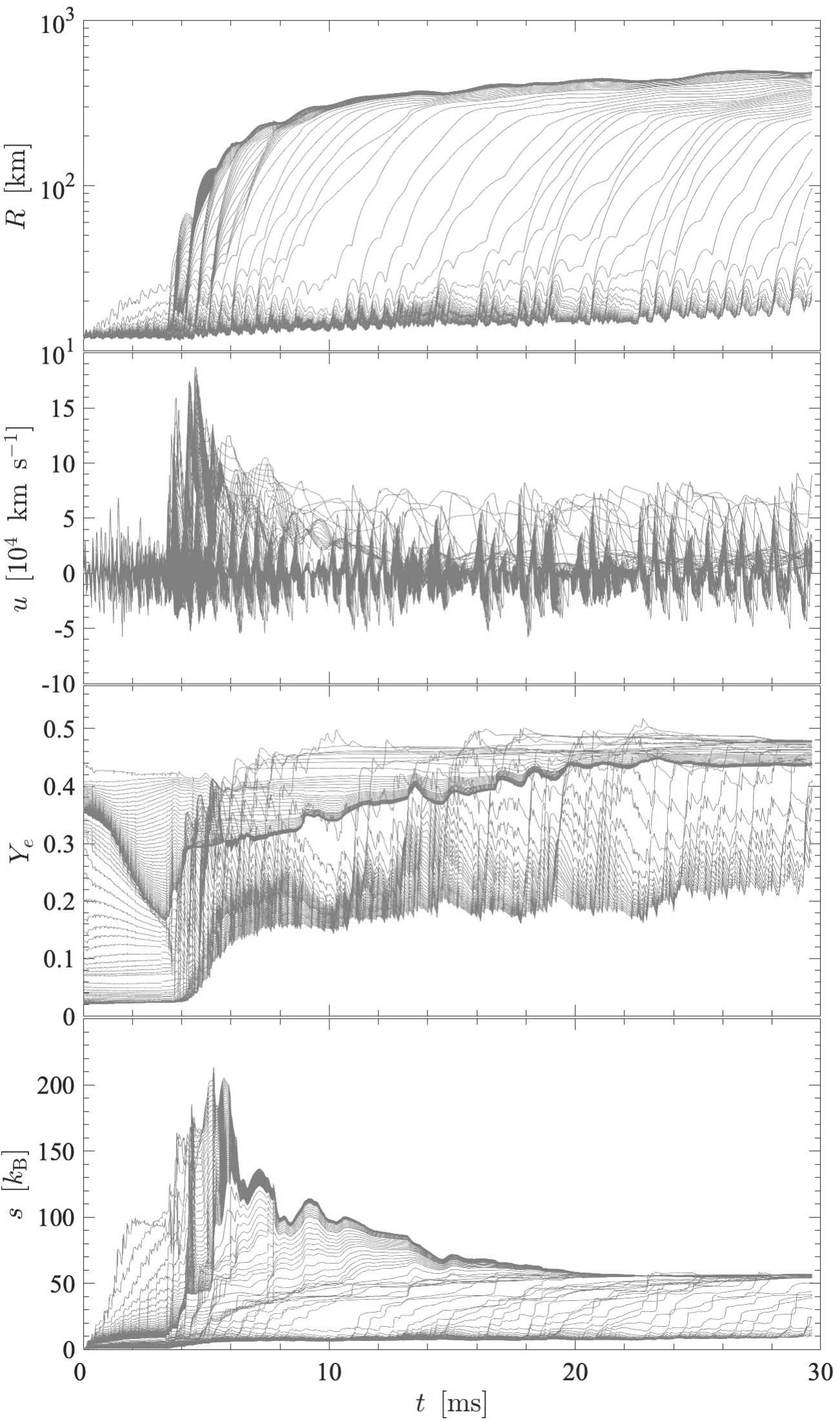}
\caption{~Evolution of 230 selected baryonic mass shells corresponding to the material being ejected. From top to bottom, the radius, $R$; velocity, $u$; electron fraction, $Y_e$; and rest-mass density, $\rho$, are shown.}
\label{fig:shellplot_RDF}
\end{figure}

\begin{figure}[t!]
\centering
\includegraphics[angle=0.,width=1.\columnwidth]{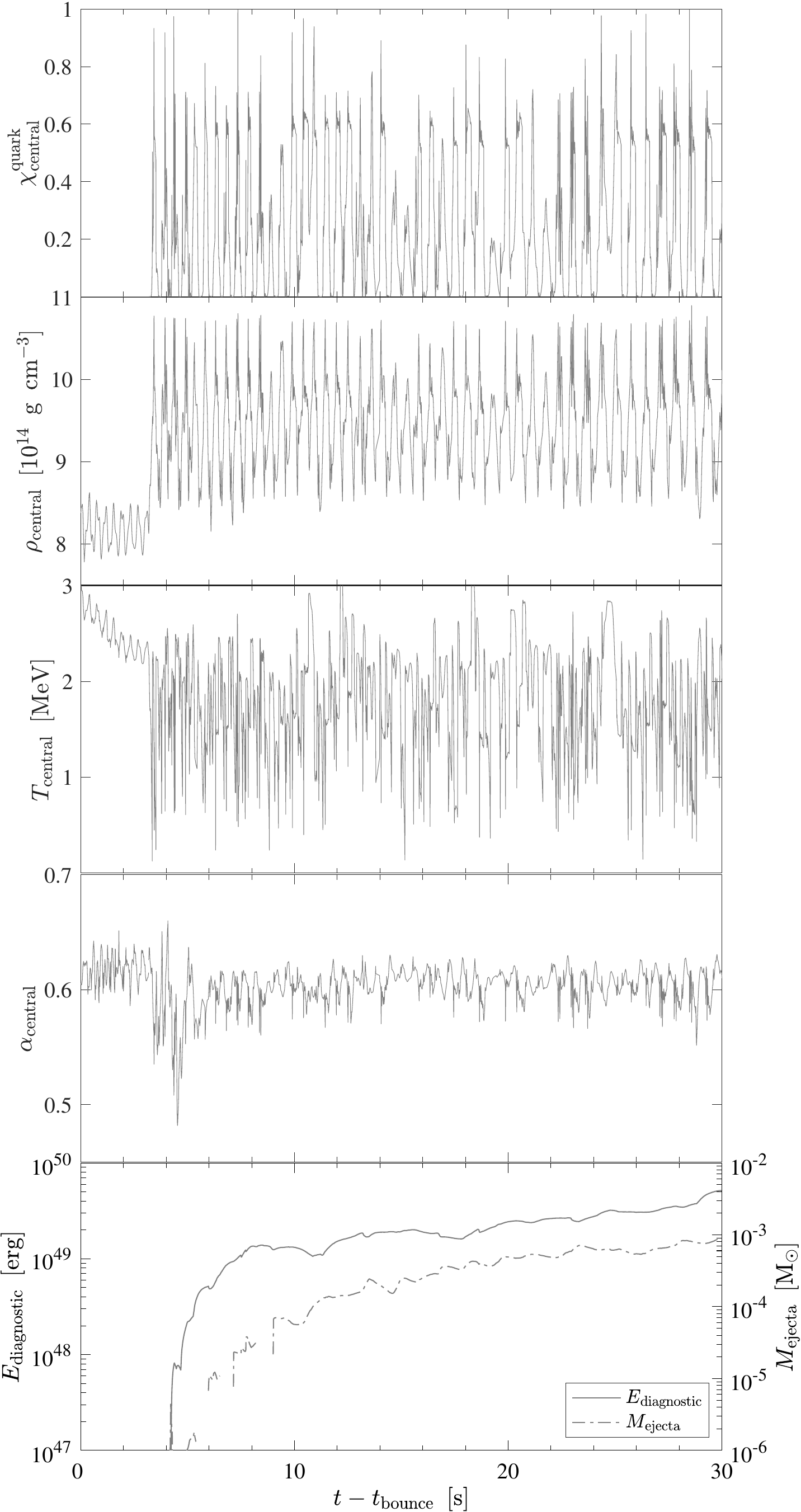}
\caption{~Evolution of selected quantities. From top to bottom, the figure shows the central quark matter volume fraction, $\chi_{\rm central}^{\rm quark}$; central rest-mass density, $\rho_{\rm central}$; central temperature, $T_{\rm central}$; central lapse function, $\alpha_{\rm central}$; and explosion energy estimate, $E_{\rm diagnostic}$ (left scale), as well as the ejected baryonic mass, $M_{\rm ejecta}$ (right scale).}
\label{fig:central}
\end{figure}

\begin{figure*}[htp]
\begin{center}
\includegraphics[angle=0.,width=1.85\columnwidth]{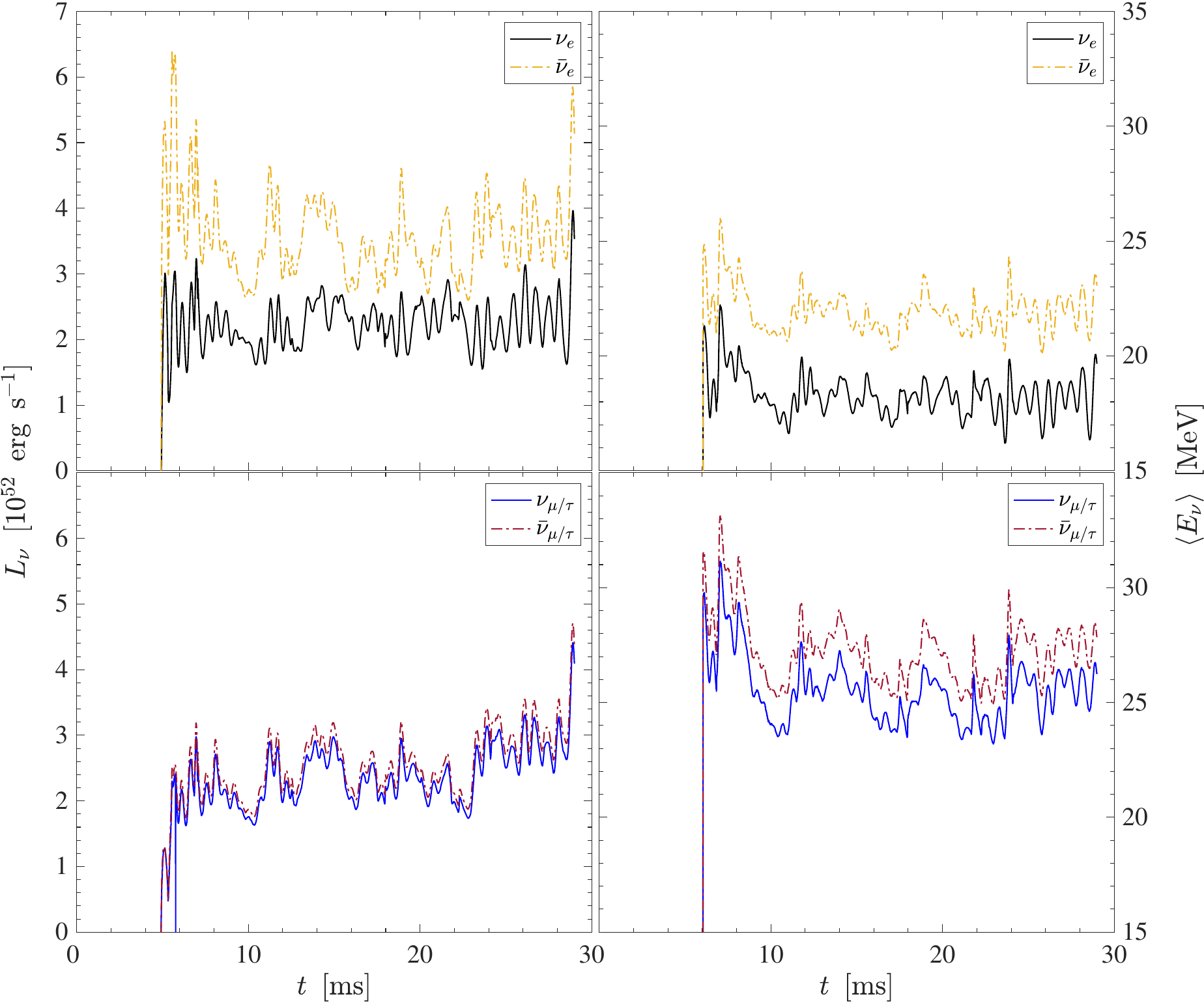}
\caption{~Evolution of the neutrino luminosities (left panels) and average energies (right panels) for all flavors sampled in the co-moving frame of reference at a radius of 200~km for the DD2F-RDF-1.1 run.}
\label{fig:neutrino_RDF}
\end{center}
\end{figure*}

We therefore employed the spherically symmetric general relativistic neutrino radiation hydrodynamics model {\tt AGILE-BOLTZTRAN}. The model was developed by \citet{Liebendorfer04ApJS150} (see also \citet{Mezzacappa93a,Mezzacappa93b,Mezzacappa93c}).
It solves the neutrino radiation hydrodynamics equations in co-moving coordinates \citep[][]{Liebendorfer01PhRvD63} and features an adaptive baryon mass mesh \citep[see][]{Liebendorfer02ApJS141,Fischer09}. 
The general relativistic three-flavor Boltzmann equation of \citet{Lindquist1966AnPhy37} was solved for the transport of neutrinos. 
The collision integral includes the list of weak reactions considered in this work \citep[see Table~1 of][]{Fischer2020PhRvC101}, which includes the Urca processes including the (inverse) neutron decay within the full kinematics approach and weak magnetism contributions, following \citet{Guo2020PhRvD102}; elastic neutrino nucleon-nucleus scattering following \citet{Bruenn85}, with the approximate inclusion of inelastic contributions and weak magnetism following \citet{horowitz02}; inelastic neutrino electron-positron scattering of \citet{Mezzacappa93c}; and neutrino pair processes. 
The neutrino pair processes consider electron-positron annihilation; nucleon-nucleon bremsstrahlung of \citet{Friman79} and \citet{hannestad98}, including the $\pi NN$ vertex corrections \citep[see also][]{Bartl2016PhRvD94,Guo2019ApJ887} implemented by \citet{Fischer16}; and the annihilation of electron neutrino-antineutrino pairs to $\mu/\tau$ neutrino-antineutrino pairs of \citet{Buras06a} implemented in \citet{Fischer09}. 

The initial neutrino phase space distributions were set to zero at all neutrino energies and propagation angles for all flavors. 
The time resolution was then adjusted to a very short scale, with time steps on the order of sub-nanoseconds, to encompass all reaction rates, including the scattering and pair reaction kernels.
This led to the launch of an artificial startup neutrino burst. However, this does not cause any impact on the evolution of the NS structure. 
This startup feature is also present in the reference hadronic run discussed in Appendix~\ref{sec:appendix1}. We discuss the later neutrino evolution and impact of neutrino transport below in the current section.

Figure~\ref{fig:shellplot_RDF} shows the evolution of 230 selected baryonic mass shells between the mass coordinates $M=1.776$~M$_\odot$ and $M=1.780676$~M$_\odot$, which corresponds to the material being ejected in the simulations launched based on the DD2F-RDF-1.1 hybrid EOS. After an initial period during which the accreted envelope expands slightly, as a direct response to the thermal pressure in the radiation-dominated low-density domain, the central quark matter fraction, denoted as $\chi^{\rm quark}$, rises slowly, as illustrated in the top panel of Fig.~\ref{fig:central}, after about 3~ms. However, the central fluid element barely reaches pure quark matter and does not remain in the pure quark matter phase. This phase is defined as $\chi^{\rm quark}=1$, where $\chi^{\rm quark}=0$ corresponds to matter in the hadronic phase and $0<\chi^{\rm quark}<1$ corresponds to matter in the hadron-quark mixed phase \citep[further details about the definitions of $\chi^{\rm quark}$ can be found in][]{Bastian2021PhRvD103}. The matter velocities, shown in Fig.~\ref{fig:shellplot_RDF}, oscillate accordingly on the order of a few times $10^4$~km~s$^{-1}$, reaching $1.5\times 10^5$~km~s$^{-1}$ at the peaks. 

After the softening of the central NS EOS due to the onset of the phase transition--- we note that the hybrid EOS softens significantly in the hadron-quark mixed phase, where the pressure slope decreases (see Fig.~\ref{fig:eos_MR})---the central density rises rapidly. 
It oscillates in accordance with the central quark matter fraction around a mean value of about $10^{15}$~g~cm$^{-3}$, as is illustrated in the top panel of Fig.~\ref{fig:central}. 
The onset of the phase transition at the NS interior is accompanied by a decreasing central temperature, as shown in Fig.~\ref{fig:central}. This phenomenon is known for this class of RDF hybrid EOSs, that is, during the adiabatic compression of the central NS fluid elements, the temperature decreases. It has been discussed in the example of core-collapse supernovae in \citet{Fischer18} and \citet{Fischer:2021}. 
In the hadron-quark mixed phase, after around 3~ms, the temperature of the central fluid elements oscillates accordingly. However, the magnitude of these oscillations is somewhat larger than before reaching the mixed phase and is due to the substantially larger EOS gradients encountered in the mixed phase. 

As a consequence of the evolution, after reaching quark matter at the NS interior, the steeper gravitational potential results in increased temperatures at the NS surface, where the accreted envelope expands and reaches supersonic matter velocities. This rapid initial expansion is illustrated in Fig.~\ref{fig:shellplot_RDF} (top panel). 
The entire envelope expands quickly to radii on the order of several 100~km. At about 15~ms, the outer envelope reaches a radius of about 400~km. 
This evolution is accompanied by a constant change in the gravitational potential, illustrated via the central lapse function $\alpha_{\rm central}$ in Fig.~\ref{fig:central}, dropping from 0.62 to slightly below 0.5. Hence, oscillations of the NS surface in turn affect the central evolution, as shown in Fig.~\ref{fig:central}. 
The associated millisecond timescale of these oscillations is related to the high central densities on the order of $10^{15}$~g~cm$^{-3}$ and the Courant-Friedrichs-Lewy (CFL) conditions for the numerical scheme solving the system of a partial differential equation. In other words, the time resolution decreases by a factor of two when the density increases from $\rho=10^{14}$~g~cm$^{-3}$ to $\rho=10^{15}$~g~cm$^{-3}$. 

At the moment of the hadron-quark phase transition, when the NS envelope expands, positive diagnostic explosion energies are obtained. 
These are evaluated by integrating the total energy, which is composed of gravitational, kinetic, and internal energies and are given by the EOS, from the surface toward the center, thereby following the standard procedure of core-collapse supernova phenomenology \citep[c.f.][and references therein]{Fischer09}. 
We obtained values of about $E_{\rm diagnostic}\simeq 5\times 10^{49}$~erg at around 30~ms of simulation time, as illustrated in Fig.~\ref{fig:central} (bottom panel). However, the asymptotic value still had not been reached. 
The total mass ejected was about $M_{\rm ejecta}\simeq 1\times 10^{-3}$~M$_\odot$ for the simulation time. 
These ejecta feature high escape velocities on the order of 10-60\% of the speed of light (see Fig.~\ref{fig:shellplot_RDF}). 

It is interesting to note that these ejecta are subject to neutrino interactions as they expand, which give rise to a broad $Y_e$ distribution (see Fig.~\ref{fig:shellplot_RDF}). 
There is an early component featuring slightly neutron-rich conditions with $Y_e\simeq 0.3-0.45$ and high entropies per baryon on the order of $s\simeq 100-175~k_{\rm B}$, as illustrated in the bottom panel of Fig.~\ref{fig:shellplot_RDF}. 
This early component is followed by an isospin symmetric component with $Y_e\simeq 0.5$ and moderately low entropies per baryon of $s\simeq 50~k_{\rm B}$. 
The neutrino luminosities and average energies are shown in Fig.~\ref{fig:neutrino_RDF} for all flavors sampled in the co-moving reference frame at a radius of 200~km. 
This corresponds to the free-streaming regime. 
Luminosities rose suddenly after about 6~ms, coinciding with when quark matter appears at the NS interior, and reached values of a few times $10^{52}$~erg~s$^{-1}$, with average energies on the order of several tens of mega-electron volts.  
Before the appearance of quark matter, the neutrino luminosities initially had values on the order of $10^{49}$~erg~s$^{-1}$ and average energies on the order of a few mega-electron volts, as is discussed and illustrated in Appendix~\ref{sec:appendix1}. 
We further note that the reason the early ejecta remained low $Y_e$ is the high expansion velocities on the order of 60\% of the speed of light, while for the later ejecta, the neutrino fluxes and their average energies are responsible for the rising $Y_e$, as has been found semi-analytically for the late-time supernova ejecta by \citet{Qian1996ApJ471}. 
The spectral and flux differences remained generally small. In particular, the difference $\langle E_{\bar\nu_e} \rangle-\langle E_{\nu_e} \rangle$ remained small (on the order of a few  mega-electron volts) despite the generally high average energies (on the order of 15-25~MeV for both $\nu_e$ and $\bar\nu_e$). However, the spectral difference is insufficient to keep the expanding material neutron-rich. 
The oscillations of the neutrino luminosities and average energies displayed in Fig.~\ref{fig:neutrino_RDF} are on the order of a few milliseconds, which is substantially longer than the oscillations of the central quantities shown in Fig.~\ref{fig:central}. In fact, they are related to sampling the neutrino quantities in the co-moving frame of reference of the fully general relativistic treatment of the neutrino radiation hydrodynamics and the Boltzmann neutrino transport at a radius of 200~km. The expanding ejecta experience fluctuations on the same timescale, which in turn are due to the deceleration and the subsequent collision with the slower moving outer ejecta, which feed back to the neutrino fluxes and spectra via the velocity terms of the transport equation.

\section{Mode analysis}
\label{sec:modes}
In this section, we report the results of an in-depth mode analysis conducted following the approach of \citet{Torres-Forne2019MNRAS482}, who employed the numerical {\tt GREAT} code. 
This code solves the eigenvalue problem of hydrodynamic and metric perturbations of a spherically symmetric self-gravitating system equilibrium model in general relativity. 
Assuming spherical symmetry, these eigenmodes are obtained through linear adiabatic perturbation analysis of the hydrodynamics and metric equations with a background that is in equilibrium. 
This linear Eulerian perturbation is expanded with a harmonic time dependence of frequency and in spherical harmonics for the angular dependence \citep[see][]{Torres-Forne2019MNRAS482}.
The corresponding Lagrangian fluid element displacement vector can thus be expressed as follows:
\begin{eqnarray}
\pmb{\xi} &=& \bigg [ \eta_r (r) Y_{lm} (\theta,\varphi) \pmb{\hat r} \nonumber \\
&& 
+\,\eta_{\perp}(r) 
\left( 
\frac{\partial_\theta Y_{lm}(\theta,\varphi)}{r^2} \pmb{\hat \theta}
+ \frac{\partial_\varphi Y_{lm}(\theta,\varphi)}{r^2 \sin^2\theta} \pmb{\hat \varphi} 
\right) 
\bigg] e^{i \sigma t},
\end{eqnarray}
where we use spherical coordinates $(r,\theta,\varphi)$ with the corresponding orthonormal basis $(\pmb{\hat r}, \pmb{\hat \theta}, \pmb{\hat \varphi})$ and the spherical harmonics $Y_{lm}$. 
Here, $\sigma$ is the oscillation frequency in the harmonic time-dependent part, whereas $\eta_r$ and $\eta_\perp$ are two functions of the radius that characterize the radial and angular components of the displacement, respectively.

\begin{figure}[t!]
\includegraphics[angle=0.,width=1.0\columnwidth]{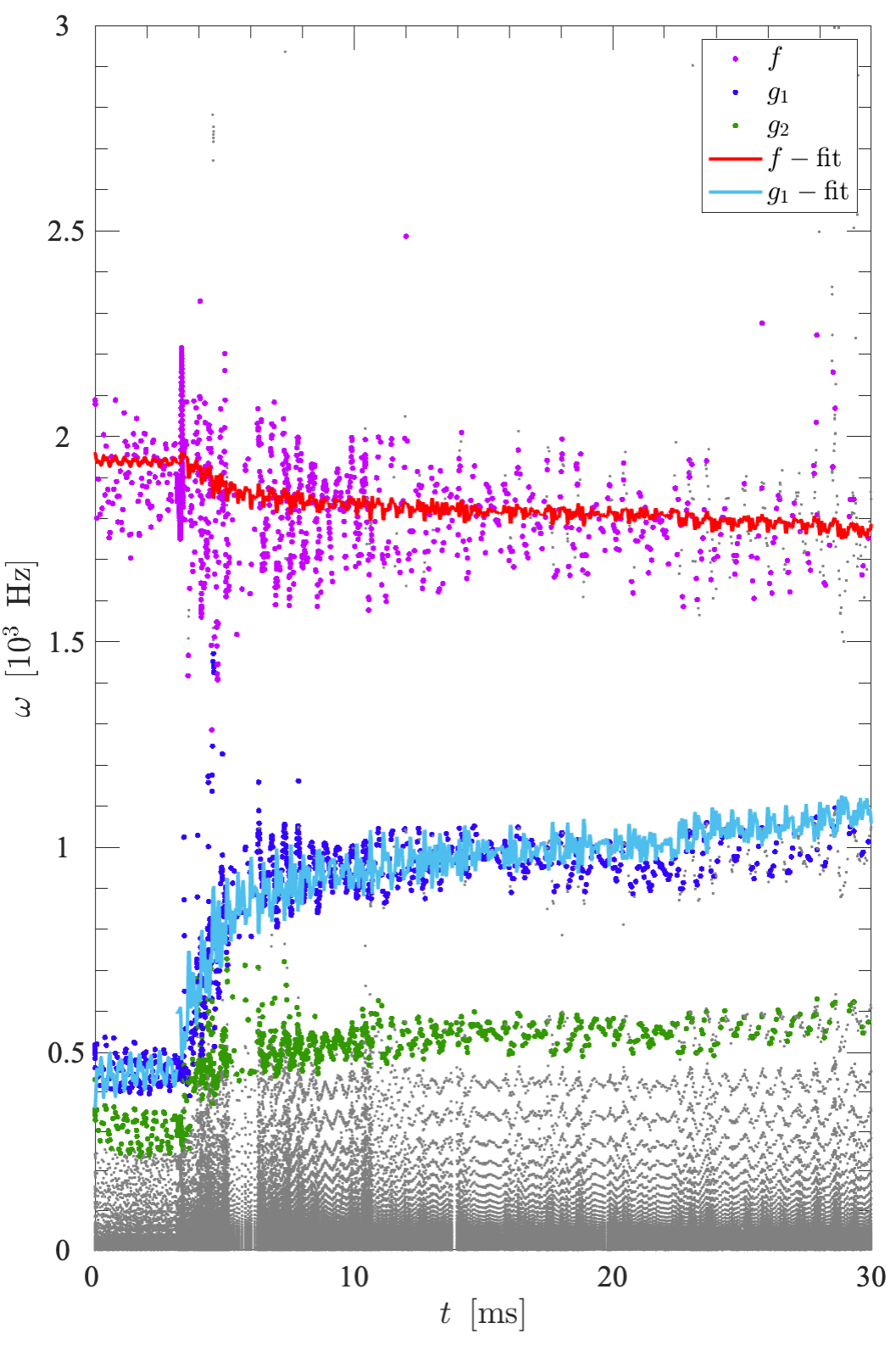}
\caption{~Results of the mode analysis showing the solutions of the {\tt GREAT} analysis ( gray dots). Highlighted in the figure are the $f$ (magenta dots), $g_1$ (dark blue dots), and $g_2$ modes (green dots) as well as the fit expressions~\eqref{eq:fit} for the $f$ (solid red line) and $g_1$ modes (light blue lines). The results were obtained with an outer boundary located at $\rho_{\rm outer}=10^{12}$~g~cm$^{-3}$, that is, excluding possible contributions from low-density solutions that might give rise to $p$ modes (see text for details).} 
\label{fig:frequencies_RDF}
\end{figure}

\begin{figure}[t!]
\includegraphics[angle=0.,width=1.\columnwidth]{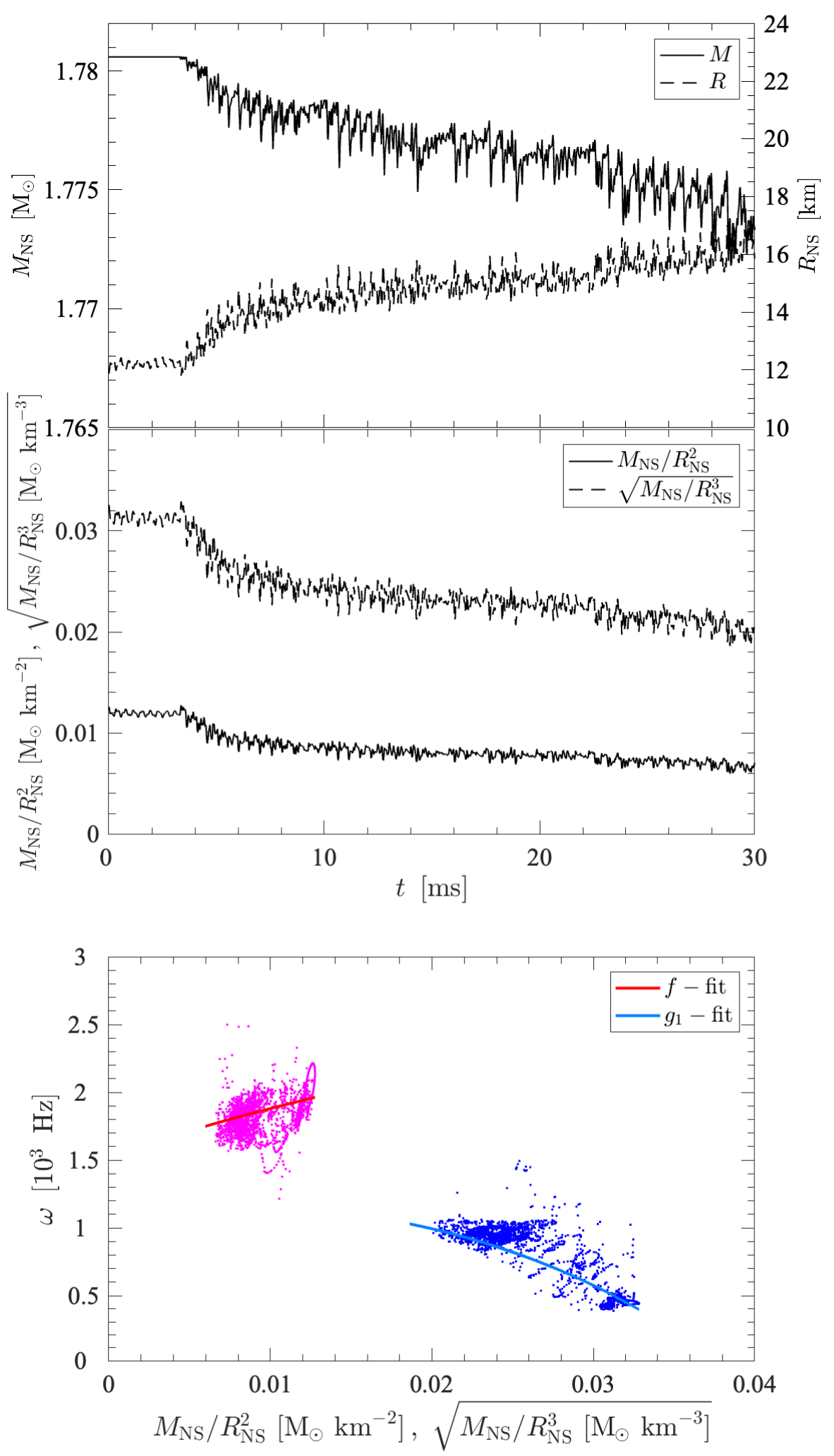}
\caption{~Evolution of selected quantities and trends of the eigenmodes. Top panels:~Evolution of the NS enclosed mass $M_{\rm NS}$ and radius $R_{\rm NS}$ sampled at a rest-mass density of $ \rho=10^{12}$~g~cm$^{-3}$. Bottom panel:~Dependencies of the $f$- and $g_1$-mode frequencies (magenta and blue dots) with respect to the mean density and surface gravity, respectively, along with the corresponding $f$- and $g_1$-mode fit expressions~\eqref{eq:fit}.}
\label{fig:MR_RDF}
\end{figure}

\begin{figure*}[htp]
\begin{center}
\includegraphics[angle=0.,width=1.925\columnwidth]{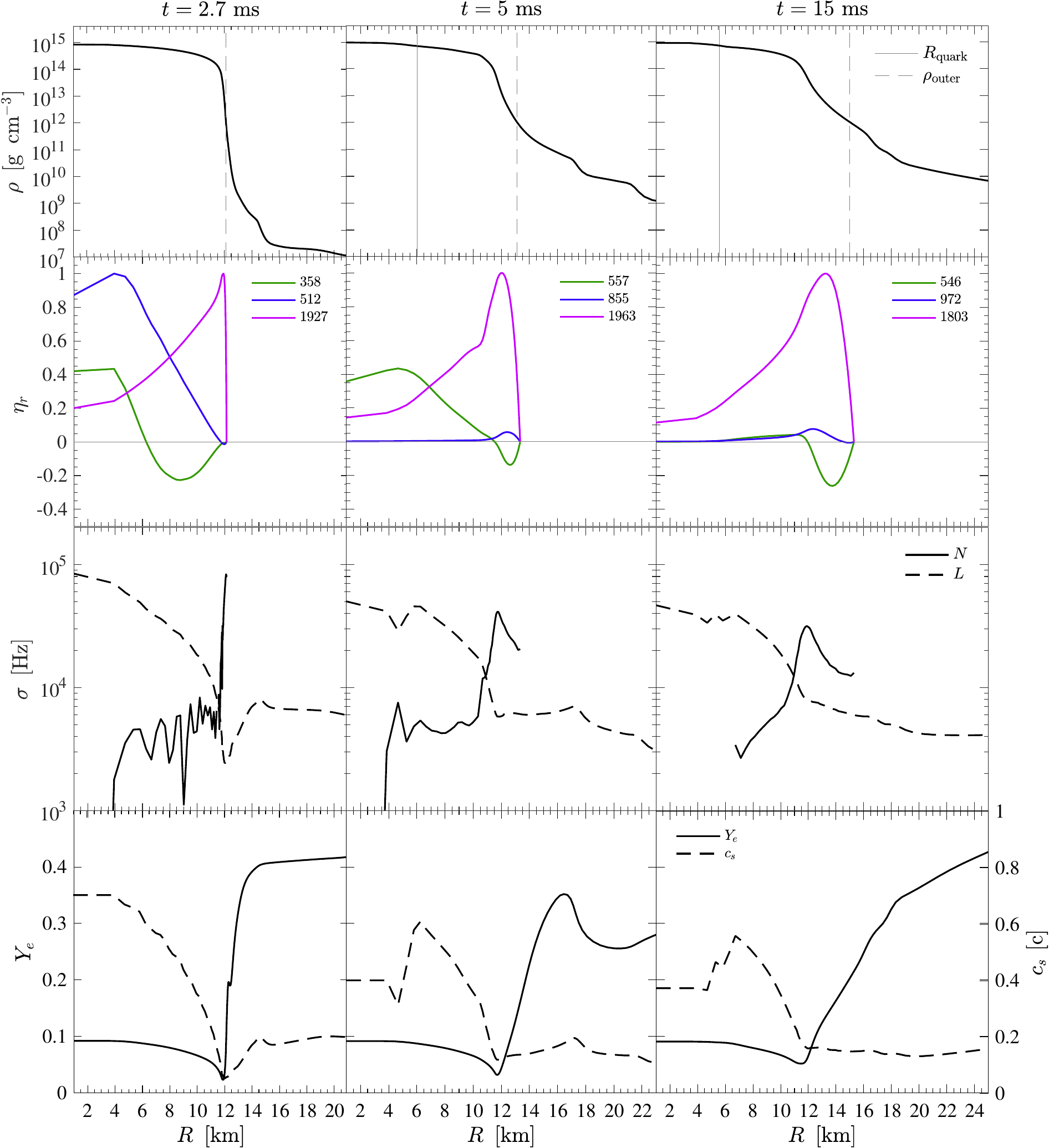}
\caption{~Radial profiles of selected quantities. From top to bottom, the figure shows the rest-mass density $\rho$; the radial solutions of the eigenfrequencies $\eta_r$ for the highest three modes found, indicating the corresponding frequencies ( values in the legends are in units of hertz and use the same colors as those in Fig.~\ref{fig:frequencies_RDF}); the Lamb $L$ and Brunt-V{\"a}is{\"a}l{\"a} $N$ frequencies as well as the $Y_e$; and the speed of sound $cs$ at three different times (see Fig.~\ref{fig:frequencies_RDF}), corresponding to the hadronic evolution in graph~(a), the transition when quark matter appears in graph~(b), and at late times in graph~(c). The vertical lines in the density plots (top panels) mark the locations of the outer boundary condition $\rho_{\rm outer}$ (dashed line) and the location for the onset of the quark matter core $R_{\rm quark}$ (solid line).}
\label{fig:modes_RDF}
\end{center}
\end{figure*}

Employing the relativistic Brunt-V\"ais\"al\"a frequency and the relativistic Lamb frequency, denoted as $N^2$ and $L^2$, respectively \citep[see Eqs.~(33) and (34) in][]{Torres-Forne2018MNRAS474}, is a valuable approach to identifying regions of possible convection. 
The former is related to the criterion of Ledoux stability and quantified by the presence of gradients of both lepton number and entropy  \citep[c.f.][and references therein]{mirizzi16}. 
In general, $N^2$ governs gravity ($g$) modes, and $L^2$ dictates pressure ($p$) modes and the fundamental ($f$) mode. 
 
Distinct modes can emerge but are contingent on the temporal evolution of diverse physical parameters. 
Gravity modes manifest in regions where buoyancy serves as a restoring force, characterized by $N^2>0$, indicating stability to convective motion. 
In the absence of buoyancy, $p$ modes are excited by the propagation of sound waves. Our focus lies not in the modes supported by sound waves but rather in the interior of the NS. 
Although the NS is not fully isolated and the choice of boundary conditions influences $p$-mode eigenfrequency calculations, we directed our attention solely toward the NS interior. 
We imposed boundary conditions from the innermost fluid elements to the surface of the NS. At $R=0$, we imposed regularity of the solution, while at the surface we imposed zero radial displacement ($\eta_r =0$) \citep[see][]{Torres-Forne2018MNRAS474}. For the computation of the $f$- and $g$-mode frequencies, it has been shown that the boundary condition used does not significantly change the frequencies \citep{sotani2019b,Wolfe2023}.
It is crucial to note that the definition of the NS surface is based on the density profile reaching $ \rho=10^{12}$~g~cm$^{-3}\equiv\rho_{\rm outer}$. 
The inner boundary was set to be the central fluid element. 
Altering the inner boundary to somewhat lower densities (e.g., $\rho=10^{14}$~g~cm$^{-3}$) has a negligible impact on the solutions given the lack of significant density fluctuations within the NS inner core. 
 In Appendix~\ref{sec:appendix2}, we present a sensitivity analysis of the mode analysis with respect to the choices of the inner and outer boundary conditions.
For further analysis, we selected as the inner boundary the density condition $\rho_{\rm inner}=10^{15}$~g~cm$^{-1}$. It corresponds to the central fluid element at $R=0$. When the central density exceeds this value, due to the oscillatory behavior of the NS evolution after the onset of quark matter at the NS interior (see Fig.~\ref{fig:central}), the boundary location is adjusted accordingly. However, as demonstrated in Appendix~\ref{sec:appendix2}, the linear mode analysis remains unchanged due to varying inner boundary conditions in the density range of $\rho_{\rm inner}=10^{13}-10^{15}$~g~cm$^{-3}$.

The frequency outcomes derived from the mode analysis via the {\tt GREAT} tool are depicted in the left panel of Fig.~\ref{fig:frequencies_RDF}. 
These results encompass solutions obtained for all nodes with $l=2$, which span the frequency range from 0 to 4~kHz. 
Due to the chosen boundary conditions, the majority of modes associated with $p$ modes fall beyond this range. Following the mode classification of \citet{Cowling1941MNRAS.101..367C}, the foremost dominant mode, characterized by zero nodes and identified as the $f$ mode, initiates at approximately 1.9~kHz. 
Subsequently, reaching the onset density for the first-order QCD phase transition, it experiences a slight reduction and transiently follows a descending trajectory. 
At lower frequencies, another prominent mode with $n = 1$ node emerges as the first $g$ mode, with a broader frequency spectrum compared to the $f$ mode and a distinctive behavior. 
It commences at approximately 0.5~kHz and exhibits a sudden, drastic rise to frequencies around 1~kHz after reaching the onset density for the QCD phase transition. 
Despite the NS never fully reaching a quark matter core and undergoing oscillations in mass and radius, the $g$ mode, on the whole, evolves toward higher frequencies.

In accordance with the node definition proposed by \citet{Torres-Forne2018MNRAS474}, the quantity $n$ is herein defined as the count of sign changes within the radial function $\eta_r$. 
These are shown in Fig~\ref{fig:modes_RDF} at three different times that correspond to the hadronic evolution before the appearance of quark matter (left panel), during the transition when quark matter appeared at the NS interior (middle panel), and during the subsequent long-term evolution (right panel). 
The associated frequencies are listed in the legend, and they coincide with the definition of \citet{Torres-Forne2018MNRAS474}, confirming the highest frequency mode to be the $f$ mode and the other two lower frequencies as $g$ modes. 
The radial profiles of both, $N$ and $L$ frequencies, are shown in Fig.~\ref{fig:modes_RDF}. 
We note that the plane-wave approximation imposes the condition $L^2>N^2$ for acceptable solutions. 
In other words, oscillatory solutions can occur when $\sigma^2>N^2, L^2$ (representing acoustic waves) or when $ \sigma^2 < N^2, L^2 $ (indicating gravity waves). 
Intermediate values lead to non-oscillatory evanescent solutions. 
It is essential to acknowledge that this simplifying assumption may not be applicable in the present context. 
The NS system features finite entropy, potentially deviating from the conditions under which this approximation is valid. 
In particular, it becomes evident from Fig.~\ref{fig:modes_RDF} that the rise of $N^2$ is dominated by the presence of a large lepton number gradient, indicated by the rising $Y_e$ profile at radii above approximately 12~km, since the lepton number and entropy gradients at the NS interior are negligible, as the initial conditions assume a constant entropy per particle at the NS interior. 
Entropy fluctuations remain small overall during the entire simulation, and the speed of sound, $c_s$, shown in the bottom panel of Fig.~\ref{fig:modes_RDF} is small, $(c_s/{\rm c})^2\simeq0.1$-$0.3$, and nearly constant toward larger radii. 
At the interior, however, the value of the speed of sound exceeds the perturbative QCD limit of $(c_s/{\rm c})^2=1/3$ since the conditions correspond to the highly non-perturbative regime of QCD. 
We further note the evolution of the $f$ and $g_1$ frequencies via the associated radial solutions $\eta_r$ in Fig.~\ref{fig:modes_RDF}, which quantifies the results shown in Fig.~\ref{fig:frequencies_RDF} at the three selected times. 
We mark the location of the outer boundary, defined as $\rho_{\rm outer}=10^{12}$~g~cm$^{-3}$, via vertical dashed line in the top panel of Fig.~\ref{fig:modes_RDF}, while the inner boundary corresponds to the very center at $R=0$. We note that the onset radius of the hadron-quark mixed phase $R_{\rm quark}$ is marked by a vertical solid line in the top panels of Fig.~\ref{fig:modes_RDF}. This onset radius is defined as the outermost location where the quark volume fraction exceeds $\chi^{\rm quark}\simeq 10^{-10}$, and it is absent in the left panels, as there has been no phase transition yet.

\begin{table}[htp]
\caption{Parameters for the frequency fits (i.e., expression~\eqref{eq:fit}).}
\begin{tabular}{c c c c c}
mode & $x$ & $a_0$ & $a_1$ & $a_2$ \\
\hline
\hline
$f$ & $\sqrt{M_{\rm NS}/R_{\rm NS}^3}$ & $1470$ & $1.5\times 10^4$ & $0$ \\
$g_1$ & $M_{\rm NS}/R_{\rm NS}^2$ & 1075 & $8\times 10^3$ & $-5.4\times 10^6$ \\
\hline
\end{tabular}
\label{tab:fit}
\end{table}

In order to fit the frequency results to the evolution of the NS bulk properties, such as the enclosed mass and its radius, we closely followed the discussion in \citet{Torres-Forne12019PhRvL123} and expressed the linear and quadratic frequency fits for the dominant and the first sub-dominant modes, respectively, in terms of the mean density inside the NS, $x\equiv\sqrt{M_{\rm NS}/R_{\rm NS}^3}$, and in terms of surface gravity, $x\equiv M_{\rm NS}/R_{\rm NS}^2$, as follows:
\begin{equation}
\omega_{\rm fit}(x) = a_0 + a_1\,x + a_2\,x^2~,
\label{eq:fit}
\end{equation}
with the fit parameters $a_0$, $a_1$, and $a_2$ are given in Table~\ref{tab:fit} for the dominant and the first sub-dominant modes identified. This indicates that from the frequency fits that are found, the dominant mode, which is initially at around 2~kHz and descends toward 1.6~kHz after the phase transition, is an $f$ mode, and the one starting at around 500~Hz and rapidly rising to a frequency of about 1000~Hz is the $g_1$ mode. The fits of the $f$ mode and $g_1$ mode are shown in Fig.~\ref{fig:frequencies_RDF} in red and light blue solid lines, respectively. These fits agree with the overall behavior of the solutions of the eigenvalue problem (i.e., the magenta and the dark blue dots, respectively). The corresponding evolution of the enclosed NS mass and radius are shown in the two top panels of Fig.~\ref{fig:MR_RDF}, sampled at a rest-mass density of $\rho=10^{12}$~g~cm$^{-3}$ to avoid $p$ modes. 
For comparison, we also show the dependency of the $f$- and $g_1$-mode frequencies with respect to the mean density and surface gravity, respectively, in the bottom panel of Fig.~\ref{fig:MR_RDF}. We include the corresponding fit expressions~\eqref{eq:fit} for comparison and demonstration of the increasing and decreasing trends with respect to these dependencies of the $f$- and $g$-modes.

The general tendency (i.e., the ratio of $M/R$) decreases after the appearance of quark matter at the NS interior. This decrease is due to the ejection of material from the NS surface, a consequence of the dynamical response of the system to the central NS oscillations discussed in Sec.~\ref{sec:simulations}. All other sub-dominant modes are higher $g_l$-modes. 

\section{Summary and conclusions}
\label{sec:summary}
The present paper discusses NSs in accreting systems as a novel astrophysical scenario for the appearance of QCD degrees of freedom. 
To this end, we performed general relativistic neutrino radiation hydrodynamics simulations based on three-flavor Boltzmann neutrino transport in spherical symmetry. 
Initial conditions were constructed based on the well-selected DD2F nuclear EOS for an NS with a total baryonic mass of 1.781~M$_\odot$, featuring an accreted envelope of $M_{\rm envelope}=0.018$~M$_\odot$, corresponding to the hybrid branch of the mass-radius relation of the DD2F-RDF-1.1 hybrid EOS employed in this study. 
This EOS features a first-order phase transition based on the Maxwell construction from the DD2F RMF nuclear matter EOS to the RDF-1.1. model. 
The latter includes vector repulsion, which is responsible for the stiffening of the EOS with increasing density and hence yields maximum hybrid star masses that are consistent with the current constraint of about 2~M$_\odot$. 
The appearance of quark matter at the NS interior causes the NS's central density to rapidly rise and oscillate on a timescale on the order of milliseconds. 
In response to the central contraction, matter is expelled from the hybrid star's surface, resulting in the ejection of approximately $10^{-3}$~M$_\odot$ of baryonic mass during the considered simulation times. 
We obtained explosion energy estimates of $5\times 10^{49}$~erg, which were computed following the procedure commonly used in the core-collapse supernova context.

Due to the sudden compression and later long-term oscillations of the hybrid star interior, which last at least for several tens of milliseconds, the hybrid star experiences a temperature increase, particularly at low densities toward the surface. 
This, in turn, results in the production of neutrinos with luminosities on the order of $10^{52}$~erg~s$^{-1}$ and average energies of several tens of mega-electron volts. 
This represents an observable signal for the current generation of kiloton water Cherenkov radiation detectors for a galactic event. 

The Urca processes, the electron captures on protons, and the positron capture on neutrons, as well as their reverse $\nu_e$ and $\bar\nu_e$ captures and the (inverse) neutron decay, cause the rise of the $Y_e$ of the ejected material from $Y_e\leq 0.46$ to $Y_e\simeq 0.5$. 
However, the release of a millisecond burst-like neutrino signature, as was reported for the QCD phase transition in the context of core-collapse supernovae \citep[c.f.][and references therein]{Fischer18}, has not been found here. 
On the other hand, a detailed GW mode analysis shows the evolution of the decreasing dominant $f$-mode frequency from around 1.9~kHz, in the hadronic phase, toward around 1.7~kHz after the appearance of quark matter. The magnitude of the decrease reported here for the particular NS mass is consistent with the findings of \citet{Sotani2023PhRvD108_universal-relations_QCD}, who further studied a range of compact stars between 1.0 and 2.3~M$_\odot$, comparing models with QCD degrees of freedom to a reference hadronic case and finding the increase of the decreasing $f$-mode frequencies for more massive stars. The present study additionally reports on the $g_1$-mode evolution, for which we find a sudden rise from 0.5~kHz, in the hadronic phase, to about 1~kHz after the appearance of quark matter. 
The lower $g_l$-mode frequencies, namely $l\geq 2$, qualitatively follow the behavior of $g_1$. 
We provided linear and quadratic fits of the $f$- and $g_1$-mode frequencies following the argumentation of \citet{Torres-Forne12019PhRvL123} in terms of the mean density and surface gravity, respectively. 

We find it interesting to note that a complete phase transition (i.e., the central NS reaches and stays in the pure quark matter) was not found even though the initial conditions belong to the hybrid branch of the mass-radius relation by an amount of exactly $M_{\rm envelope}$. 
Mass transfer in binary systems---from a main-sequence star to a secondary NS---results in the slow accumulation of material at the NS surface and hence the growth of the NS total mass. 
A low-mass NS born in a core-collapse supernova explosion without a QCD phase transition will move slowly along the mass-radius relation toward increasing mass. 
The present investigation demonstrates quantitatively for a particular hadron-quark hybrid EOS the hydrodynamical evolution when such a system reaches the conditions for the onset of the QCD phase transition. 
Notably, what remains to be explored is the role of the EOS in terms of varying onset densities for quark matter as well as bulk properties of the phase transition, such as the pressure slope and other dependencies of the hadron-quark mixed phase, including the possible presence of shear \citep[][]{Sotani2013NuPhA906_QCD-mixed-phase_shear} and the associated presence of geometric structures collectively known as "pasta" phases \citep[c.f.][and references therein]{Yasutake2014PhRvC89}. We leave these topics for future exploration. 

Furthermore, the impact of potentially high magnetic fields and the rapid rotation associated with the very existence of magnetars require the multi-dimensional magneto neutrino radiation hydrodynamics framework. 
The present work is limited to NSs with low surface magnetic fields, referred to as old NSs. 
However, the response of the system reported here due to the QCD phase transition is expected to remain qualitatively independent of the magnetic field, featuring the ejection of material from the NS surface and the rise of the neutrino fluxes to observable values as well as the emission of GW with kilohertz frequencies. 
The magnitude of the frequencies might also alter the amount of mass ejected, as the oscillations reported here will be damped by the emission of GW in multi-dimensional simulations, as was demonstrated in the multi-dimensional hydrodynamics simulations of \citet{Dimmelmeier09} based on the polytropic parametrized and MIT bag model EOS, which settles the NS back into the initial stable state on a longer timescale. 

\section*{Acknowledgements}
The authors acknowledge support from the Polish National Science Centre (NCN) under grant numbers 2020/37/B/ST9/00691 (N.K.L., T.F., S.S.) and 2019/33/B/ST9/03059 (S.S).
S.S. acknowledges support from the program Excellence Initiative--Research University of the University of Wroc{\l}aw of the Ministry of Education and Science and the Scultetus Visiting Scientist Program of the Center for Advanced Systems Understanding.
P.C.D and A.T.F acknowledge support from the Spanish Agencia Estatal de Investigaci\'on (Grants No. PGC2018-095984-B-I00 and PID2021-125485NB-C21) funded by MCIN/AEI/10.13039/501100011033 and ERDF A way of making Europe, by the Generalitat Valenciana (PROMETEO/2019/071), and by COST Actions CA16104 and CA16214.
All simulations were performed at the Wroclaw Centre for Scientific Computing and Networking (WCSS). 

\bibliographystyle{aa}
\bibliography{references}

\clearpage

\begin{appendix}
    
\section{Hydrostatic equilibrium solution}
\label{sec:appendix1}

\begin{figure}[t!]
\centering
\includegraphics[angle=0.,width=1.\columnwidth]{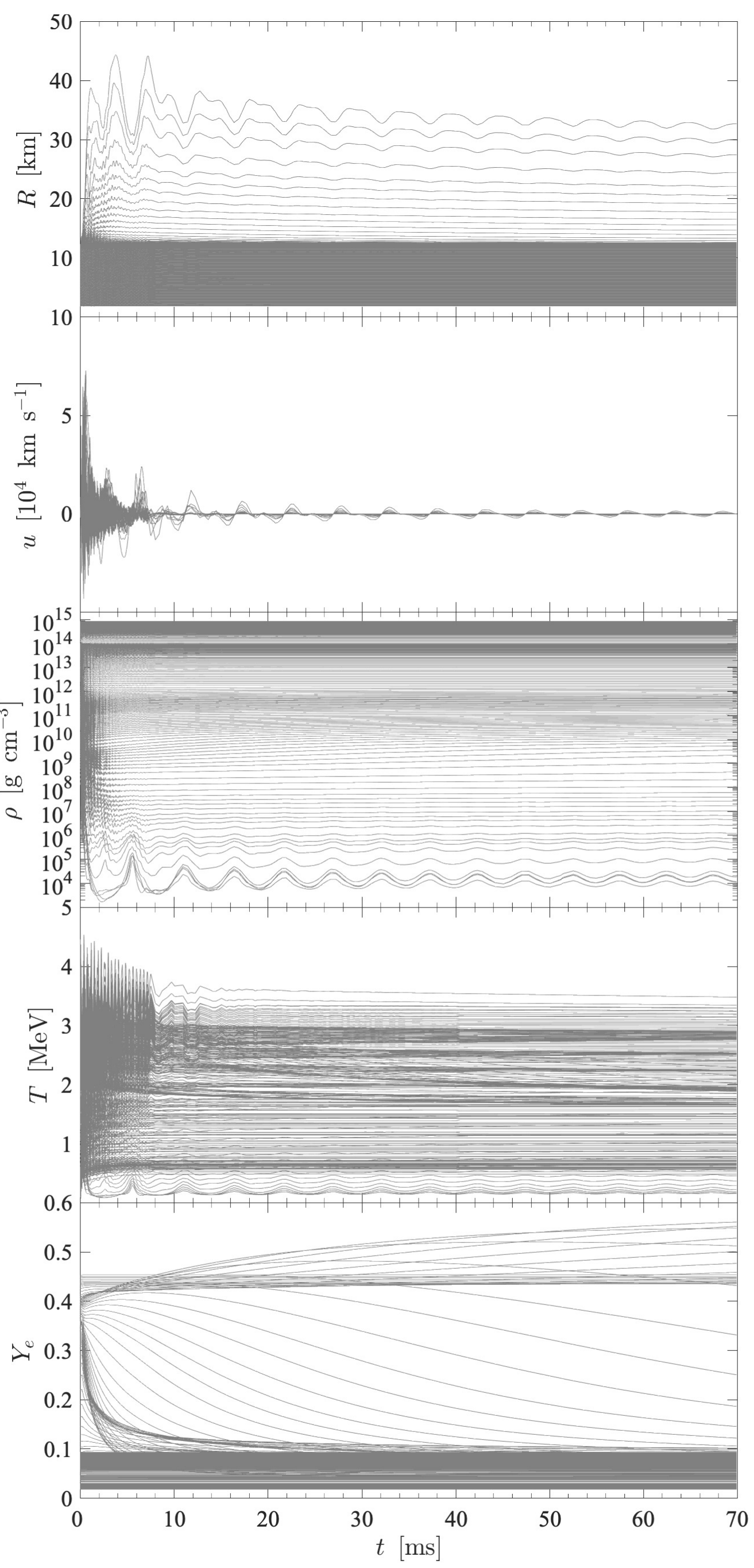}
\caption{~Evolution of mass shells for the reference DD2F hadronic model showing the evolution of (from top to bottom) radius, velocity, rest-mass density, temperature, and electron fraction.}
\label{fig:appendix_shellplot}
\end{figure}

To quantify the hydrostatic equilibrium, we launched neutrino radiation hydrodynamics simulations for the reference DD2F hadronic EOS, featuring a high radial resolution of 300 radial mass shells implemented in {\tt AGILE-BOLTZTRAN}. Otherwise, the same input physics was used as discussed in Sec.~\ref{sec:simulations}. After an initial relaxation phase that lasted for about 5-10~ms, the central compact NS settled into complete hydrostatic equilibrium. The results are illustrated in Fig.~\ref{fig:appendix_shellplot} for all 300 mass shells, showing the evolution of selected quantities, including  (from top to bottom) radius, velocity, rest-mass density, temperature, and $Y_e$. The evolution of all central mass shells remains at a constant value for all quantities and corresponds to the NS interior for densities as low as $\rho\simeq 10^{6}$~g~cm$^{-3}$. The accreted envelope, on the other hand, does not obey complete hydrostatic equilibrium, as illustrated via the expanding and contracting mass shells that reach up to radii of about 40~km and the velocities that correspondingly reach values of a few times $10^4$~km/s. Densities and temperatures follow accordingly, oscillating around $\rho\simeq 10^4$~g~cm$^{-3}$ and $T\simeq0.2$~MeV. The oscillations are damped toward later times on the order of several tens of milliseconds. 

\begin{figure}[t!]
\centering
\includegraphics[angle=0.,width=0.95\columnwidth]{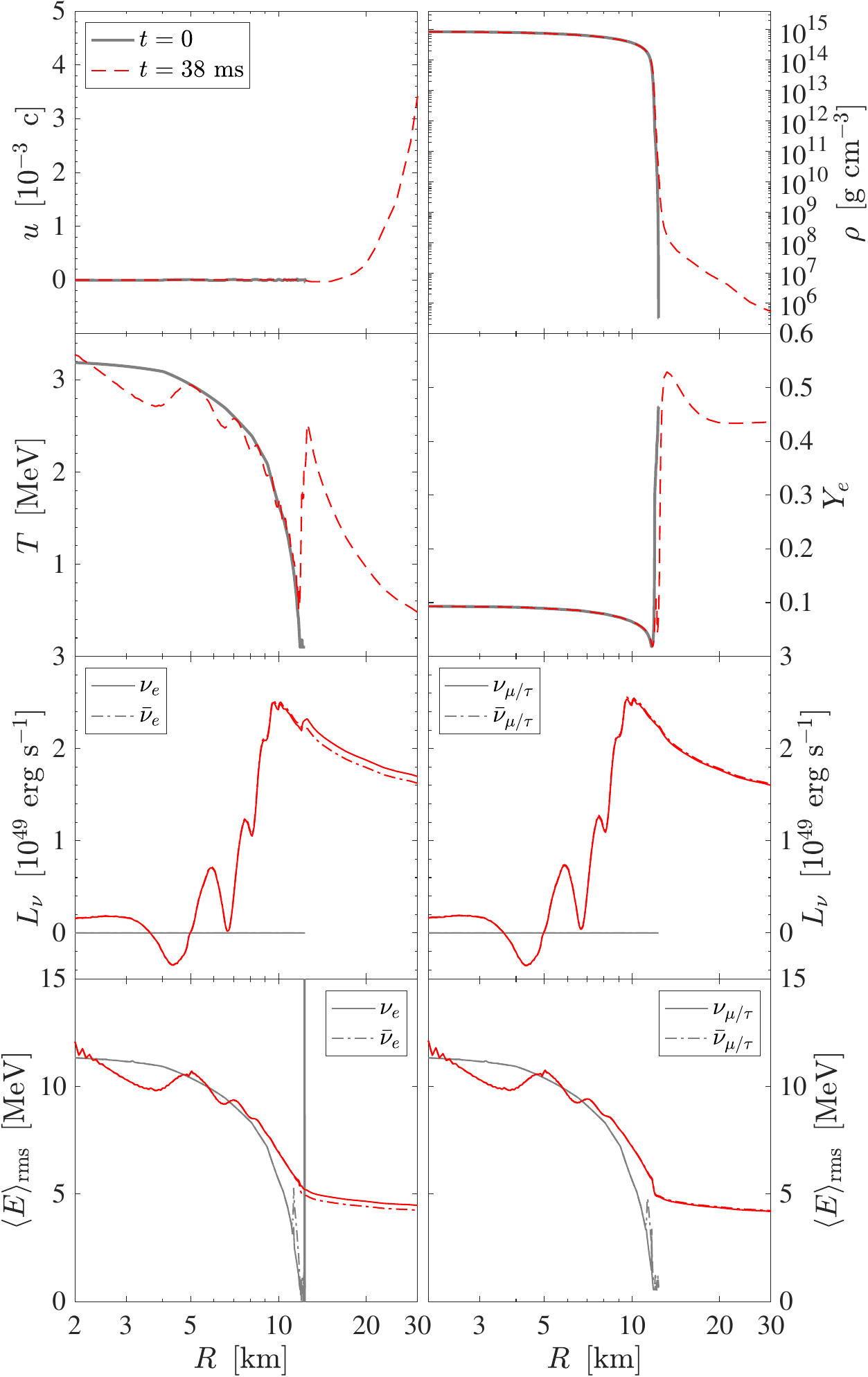}
\caption{~Radial profiles of selected quantities showing the reference DD2F hadronic simulation at $t=0$ and $t\simeq 38$~ms of simulation time. The selected quantities are velocity $u$, in units of the speed of light ${\rm c}$; rest-mass density $\rho$; temperature $T$; electron abundance $Y_e$; neutrino luminosities $L_\nu$; and root mean square energies $\langle E\rangle_{\rm rms}$ of all flavors.}
\label{fig:appendix_fullstate}
\end{figure}

All matter remains gravitationally bound at all times during the simulation, and hence no positive explosion energy estimate was obtained. The situation is illustrated via the radial profiles of selected quantities in Fig.~\ref{fig:appendix_fullstate}, which are compared to the initial conditions (thick grey lines) and the conditions at about 38~ms of the simulation time (thin red lines). The figure shows the expansion of the envelope to radii of about 100~km and the subsequent density, temperature, and $Y_e$ profiles. 

\begin{figure}[b!]
\centering
\includegraphics[angle=0.,width=1\columnwidth]{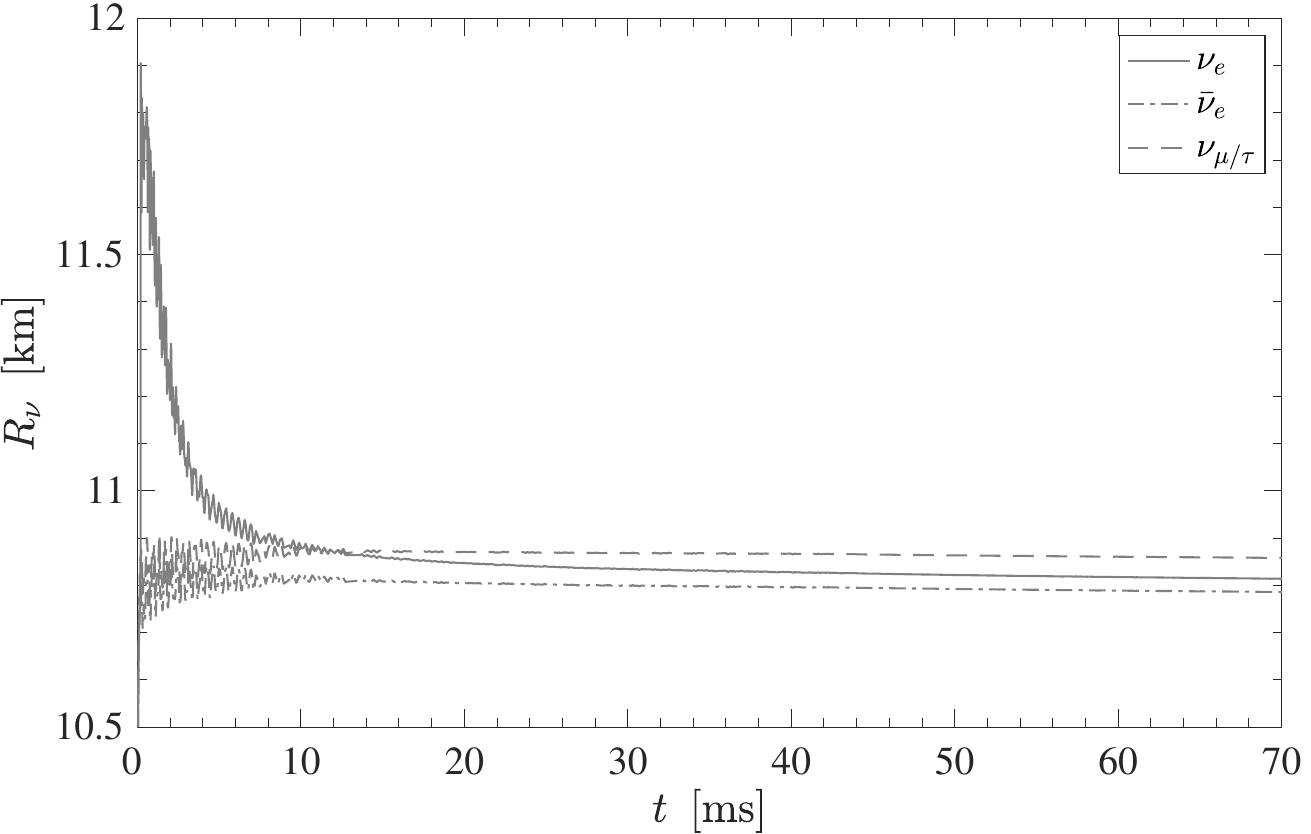}
\caption{~Evolution of the neutrinospheres $R_\nu$ of all flavors.}
\label{fig:appendix_neutrinospheres}
\end{figure}

\begin{figure}[t!]
\centering
\includegraphics[angle=0.,width=1.\columnwidth]{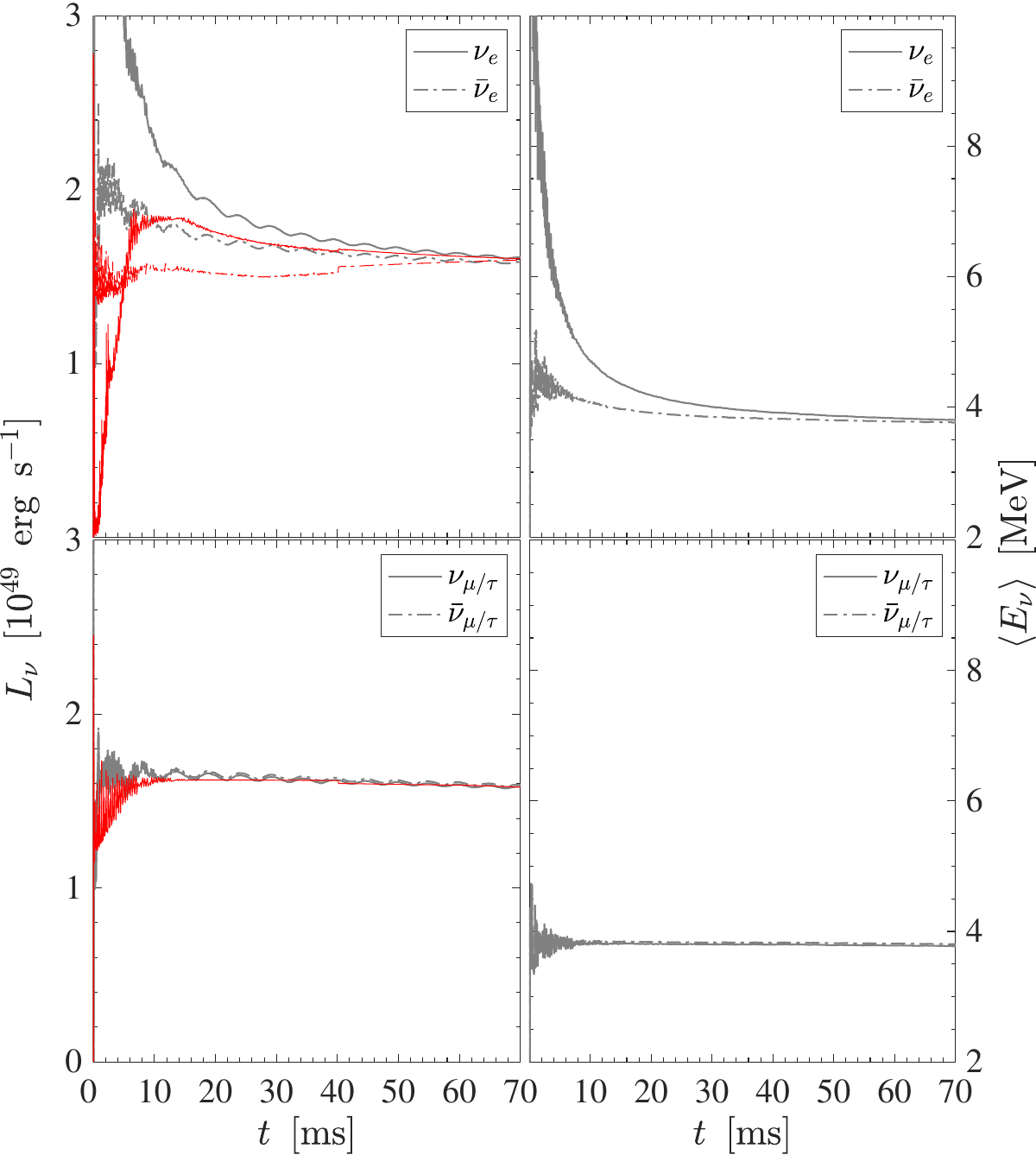}
\caption{~Evolution of neutrino luminosities (left panels) and average energies (right panels) for all $\nu_e$ and $\bar\nu_e$ (top panels) and heavy lepton flavors (bottom panels), for the DD2F hadronic reference simulations (grey lines), in comparison to the diffusion approximation (red lines) following Expression~\eqref{eq:appendix_LT}.}
\label{fig:appendix_neutrino}
\end{figure}

The evolution of the neutrino luminosities and average energies are shown in Fig.~\ref{fig:appendix_neutrino} for all flavors. The values of the luminosities remain generally low (on the order of a few $10^{49}$~erg~s$^{-1}$ and with average energies of $\langle E_\nu \rangle\simeq 4$~MeV) after the initial relaxation phase of about 10~ms. The neutrinos stem from the NS's surface, where the neutrino luminosities rise. Values of the neutrinoshere radii, $R_\nu$, are given in Fig.~\ref{fig:appendix_neutrinospheres} following the standard definition via the optical depth $\tau_\nu$, where $R_\nu:= r(\tau_\nu\equiv 2/3)$, which settles toward constant values between $R_{\nu_{\mu/\tau}}\simeq 10.75$~km, $R_{\nu_e}\simeq 10.81$, and $ R_{\bar\nu_e}=10.85$ after the initial relaxation period. The radial profile of the luminosities and root mean square energies $\langle E \rangle_{\rm rms}$ are illustrated in the bottom panels of Fig.~\ref{fig:appendix_fullstate}, which shows the transition between the NS crust and the previously accreted envelope and corresponds to the sharp rise of the envelope temperature. 

The Urca processes drive the electron abundance of the expanding envelope initially from $Y_e\simeq0.46$ toward $Y_e\simeq 0.5$ and even above at later times, as shown in Fig.~\ref{fig:appendix_fullstate}. The increasing $Y_e$ of the crust-envelope interface is also shown in the evolution of the mass shells in the bottom panel of Fig.~\ref{fig:appendix_shellplot}. This is the reason for the hierarchy of the neutrino luminosities, $L_{\nu_e}\gtrsim L_{\bar\nu_e}$, and average energies, $\langle E_{\nu_e} \rangle\gtrsim \langle E_{\bar\nu_e} \rangle$, but is nevertheless comparable to other weak processes, which are the cause of the heavy lepton flavor neutrino luminosities and the average energies being of similar order as those of the electron flavors. We note that the neutrino luminosities asymptotically approach the diffusion limit,
\begin{equation}
L_{\rm T} \simeq 4\pi \sigma_{\rm SB}\,\kappa_\nu\,R_\nu^2\,T_\nu^4~,
\label{eq:appendix_LT}
\end{equation}
where $\kappa_\nu$ is an empirical factor that is on the order of $20\%$ ($1/6$ for $\nu_e$, $10/52$ for $\bar\nu_e$, and $10/46$ for the heavy lepton neutrino flavors). The values of the neutrino luminosities~\eqref{eq:appendix_LT} are shown in Fig.~\ref{fig:appendix_neutrino} as red lines for comparison. This approximation has been explored in the core-collapse supernova case for both the fail branch in \citet{Fischer09} and the proto-NS deleptnoization as well as later cooling phases in \citet{Fischer12}. Here, particularly during the early evolution, \eqref{eq:appendix_LT} reproduced the behavior of the heavy lepton neutrino flavor. The evolution of the neutrinospheres of the last scattering is shown in Fig.~\ref{fig:appendix_neutrinospheres}, for which the method based on the integration of the transport mean free path of \citet{Fischer12} was employed. 
\FloatBarrier

\section{Role of the boundary conditions for the linear mode analysis}
\label{sec:appendix2}
A sensitivity analysis was performed on the solutions of the linear mode analysis of Sec.~\ref{sec:modes} depending on the choice of the inner and outer boundaries. Locations were defined through the rest-mass densities $\rho_{\rm inner}$ for the inner boundary and $\rho_{\rm outer}$ for the outer boundary. 

\begin{figure}[htp]
\begin{center}
\subfigure[~Frequencies.]{\includegraphics[angle=0.,width=1.0\columnwidth]{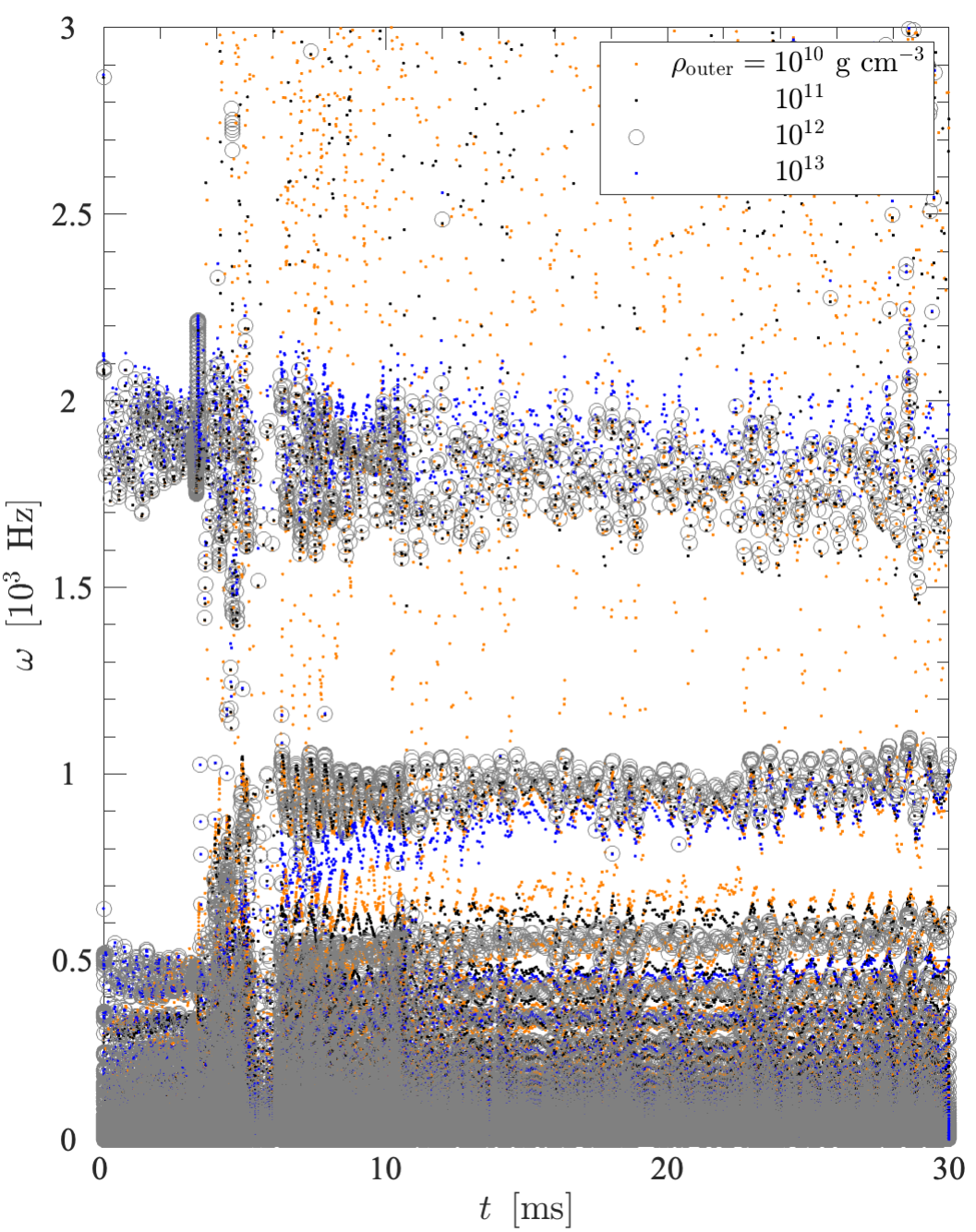}
\label{fig:frequencies_RDF_appendix_a}}
\\
\subfigure[~Neutron star mass $M_{\rm NS }$ and radius $R_{\rm NS }$ evolution.]{\includegraphics[angle=0.,width=1.0\columnwidth]{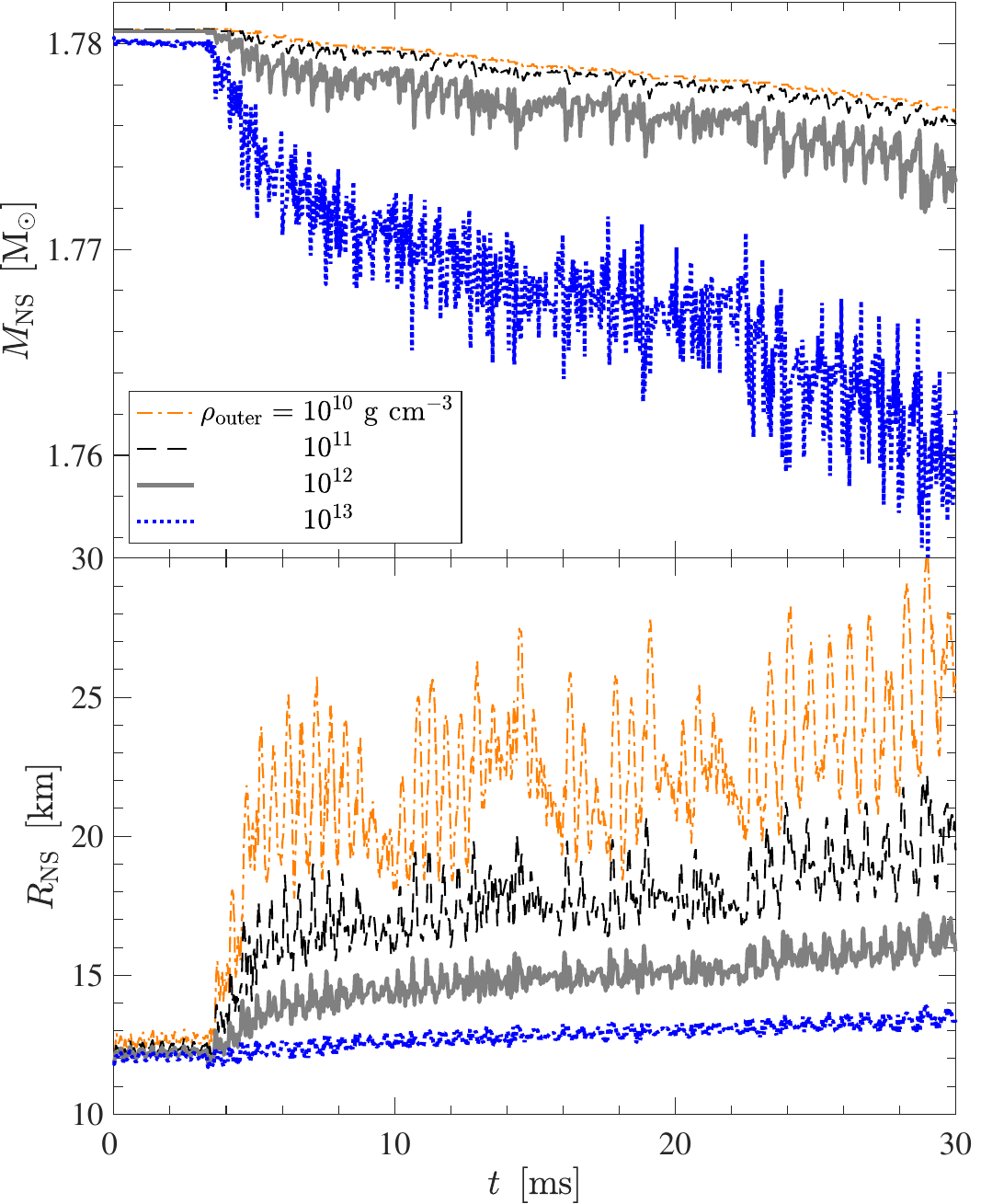}
\label{fig:MR_RDF_appendix_a}}
\caption{~Same as Fig.~\ref{fig:frequencies_RDF} but for a fixed inner boundary at $\rho_{\rm inner}=10^{15}$~g~cm$^{-3}$ and varying outer boundary conditions $\rho_{\rm outer}$.}
\label{fig:frequencies_MR_RDF_appendix_a}
\end{center}
\end{figure}

\begin{figure}[htp]
\begin{center}
\includegraphics[angle=0.,width=1.0\columnwidth]{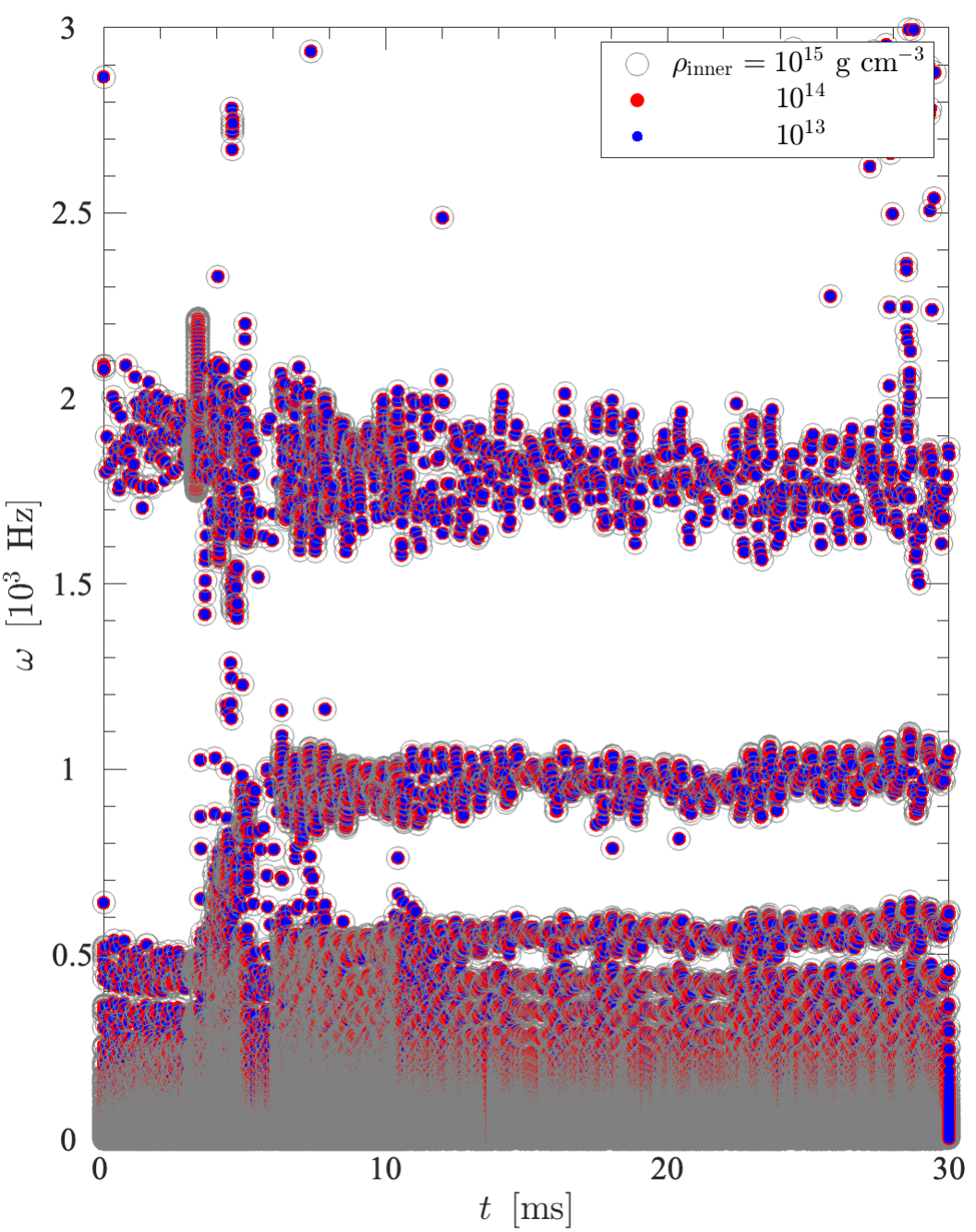}
\caption{~~Same as Fig.~\ref{fig:frequencies_MR_RDF_appendix_a} but for a fixed outer boundary at $\rho_{\rm outer}=10^{12}$~g~cm$^{-3}$ and varying outer boundary conditions $\rho_{\rm inner}$.}
\label{fig:frequencies_RDF_appendix_b}
\end{center}
\end{figure}

We first varied the outer boundary and kept the inner boundary fixed at $\rho_{\rm inner}=10^{15}$~g~cm$^{-3}$, which corresponds to the value used for the analysis presented in Sec.~\ref{sec:modes}. The results are shown in the top panel of Fig.~\ref{fig:frequencies_MR_RDF_appendix_a}. We note that when the central density should be below $\rho_{\rm inner}$, we selected the innermost grid point at $r=0$ as the inner boundary. The first observation is the reduction of the pressure modes when shifting the outer boundary to higher densities. For $\rho_{\rm outer}=10^{10}$~g~cm$^{-3}$ (orange dots), there are plenty of $p$-mode solutions present, which reduces significantly when $\rho_{\rm outer}=10^{11}$~g~cm$^{-3}$ (black dots). At $\rho_{\rm outer}=10^{12}$~g~cm$^{-3}$ (open gray circles), all the $p$-modes have disappeared. This is the setup used for the analysis performed in Sec.~\ref{sec:modes} since this work focuses on the $f$ and $g$ modes, that is, we neglect all $p$-mode contributions. The second observation is that the $f$- and $g$-mode results remain quantitatively the same for $\rho_{\rm outer}=10^{10}-10^{12}$~g~cm$^{-3}$. Only for $\rho_{\rm outer}=10^{13}$~g~cm$^{-3}$ (blue dots) do the $f$ modes shift to generally higher values and the $g$ modes shift to somewhat lower values, which is confirmed by the lower enclosed NS baryonic mass and the smaller radius, shown in the bottom panel of Fig.~\ref{fig:frequencies_MR_RDF_appendix_a} in a comparison to the other outer boundaries.  
This analysis demonstrates the sensitivity of the mode analysis to the outer boundary density range. In other words, outer boundaries above of around $\rho_{\rm outer}\simeq 10^{13}$~g~cm$^{-3}$ artificially remove $f$-mode and $g$-mode results. 

Second, we varied the inner boundary, $\rho_{\rm inner}$, while keeping the outer boundary fixed at $\rho_{\rm outer}=10^{12}$~g~cm$^{-3}$; that is, we ignored all possible $p$ modes. The results are shown in Fig.~\ref{fig:frequencies_RDF_appendix_b}, where it becomes evident that when varying the inner boundary in the range of $\rho_{\rm inner}=10^{13}-10^{15}$~g~cm$^{-3}$, the results of the mode analysis remain quantitatively the same. In fact, the different solutions of the linear system are indistinguishable, which is why we varied the marker size and color of the data shown in Fig.~\ref{fig:frequencies_RDF_appendix_b}, which all lay on top of each other. This overlap demonstrates the robustness of the results with respect to the choice of the inner boundary.

\end{appendix}
\end{document}